\documentclass[letterpaper]{article} 
\usepackage{aaai2026}  
\usepackage{times}  
\usepackage{helvet}  
\usepackage{courier}  
\usepackage[hyphens]{url}  
\usepackage{graphicx} 
\usepackage{natbib}  
\usepackage{caption} 
\usepackage{algorithm}
\usepackage{algorithmic}

\usepackage{booktabs}
\usepackage{amssymb}
\usepackage{amsmath}
\usepackage{newfloat}
\usepackage{listings}
\DeclareCaptionStyle{ruled}{labelfont=normalfont,labelsep=colon,strut=off} 
\floatstyle{ruled}
\newfloat{listing}{tb}{lst}{}
\floatname{listing}{Listing}
\title{Towards an Evaluation Methodology for AI in Second Language Education: Lessons Learned from Developing L2-Bench}
\author{
    James Edgell\textsuperscript{\rm 1}\equalcontrib,
    Wm. Matthew Kennedy\textsuperscript{\rm 2}\equalcontrib \\
    Isaac Pattis\textsuperscript{\rm 1}, Ben Knight\textsuperscript{\rm 1}, Danielle Carvalho\textsuperscript{\rm 1},     Elizabeth Wonnacott\textsuperscript{\rm 3}
}
\affiliations{
    \textsuperscript{\rm 1}Oxford University Press\\
    \textsuperscript{\rm 2}Oxford Internet Institute, University of Oxford\\
    \textsuperscript{\rm 3}Department of Education, University of Oxford

    Correspondence to: L2-Bench Team (elt-bench@oup.com)
}

\usepackage{bibentry}

\begin{document}

\nocopyright

\maketitle

\begin{abstract}
The rapid adoption of large language models in AI-powered language education has created an urgent need for evaluations that assess pedagogical effectiveness, particularly in language learning—one of the most common LLM use cases \cite{Tamkin2024Clio, costagomes2025copilot}. With only narrowly defined task-specific evaluations of AI system capabilities in second language (L2) education existing in the literature, we require more holistic approaches in this AI for education space. To address this gap, we describe the iteration of the methodology we developed to build L2-Bench, a novel, context-specific evaluation benchmark grounded in a validated “language learning experience designer” construct to assess AI capabilities across L2 education contexts. Our methodology integrates pedagogical theory, sociotechnical AI evaluation methods, and operationalizes a hierarchical taxonomy to structure an expert-curated dataset of over 1,000 authentic rubric-scored task-response pairs with measurement and scoring pipeline. We report the results of a pilot validation exercise (N = 39) on an initial sample of our dataset (tasks were validated as authentic [M = 4.23/5], but criteria scores were lower [M = 3.94], with universally poor inter-annotator agreement despite good internal consistency), alongside the experimental design for our follow-up practitioner data validation study as we iterate and scale to the full dataset. Ultimately, this research not only offers methodological lessons towards a more context-specific AI evaluations ecosystem, but also works towards better design of reproducible evaluations for AI systems deployed to educational contexts. 
\end{abstract}


\section{Introduction}

Access to language education is a human right \cite{udlr1996}. Large Language Models (LLMs) are rapidly being integrated into language learning products used by millions of learners worldwide. However, no widely recognised benchmarks exist for evaluating AI capabilities in education beyond narrow lesson generation \cite{clark-etal-2020-tydi} or knowledge of pedagogical concepts \cite{lelièvre2025benchmarkingpedagogicalknowledgelarge}. Although important aspects, learning experience design requires knowing when and how to use these capabilities in practice, reflecting the diverse educational scenarios encountered in real-world settings.

The absence of comprehensive evaluation frameworks for AI in education creates several critical challenges. Educators and institutions lack systematic methods to assess AI capabilities for specific teaching scenarios, leading to uninformed adoption decisions. Product developers cannot rigorously validate their AI implementations against pedagogical best practices. Most importantly, the lack of standardized evaluation impedes the development of more effective AI-powered educational systems.

In this paper, we present consequential  lessons learned in the construction of L2-Bench, an evaluation benchmark assessing AI performance in second language learning experience design. We discuss our iterative efforts to develop and validate artefacts produced by our methods, and in doing so, we make three contributions: 

\begin{enumerate}
    \item \textbf{A novel methodology} for developing AI evaluation benchmarks specific to the unique norms, values, and dynamics of educational spaces.
    \item \textbf{A hierarchical taxonomy} of 12 competencies and 31 subcompetencies that comprise a L2 learning experience design construct, grounded in established pedagogical framework
    \item \textbf{Lessons learned in the development of our novel methodology}, including positive and negative results of our efforts to validate our core technical components: taxonomy, measures, and an initial sample (325 items) of our planned  1,300 item dataset.
\end{enumerate}

Overall, we hope that our contributions help move AIED evaluations in general beyond narrow accuracy metrics toward more rigorous and context-sensitive assessment of AI capabilities in educational spaces. 

\section{Related work}
The speed and scale of AI system deployment have outpaced the development of evaluation methodologies for novel technologies and real-world contexts \cite{10.5555/3716662.3716698, bean2025measuringmattersconstructvalidity, reuel2024betterbenchassessingaibenchmarks, schwartz2025realitychecknewevaluation}, prompting calls for greater rigor, systematization, and the incorporation of social scientific methods \cite{butler2024futureofwork, weidinger2025evaluationsciencegenerativeai, olteanu2025rigoraidoingrigorous}. 

This evaluations crisis is acute in education, where no generally accepted holistic evaluations for AI in education exist despite rapid adoption \cite{digitaleducationcouncil2024globalai, costagomes2025copilot}. While some evaluations assess performance in highly rules-based domains (e.g., mathematics or computer science), these are not readily transferable to more open-ended educational settings. Conversely, benchmarks claiming to measure broad constructs such as knowledge or reasoning risk importing inappropriate measurement techniques into educational contexts. These limitations create conditions for substantial, compounding risk \cite{Bastani2024HarmLearning, Kennedy2026Vernacularized}. 

Evaluations for AI in education are only beginning to emerge. A growing body of scholarship suggests AI use may yield marginal learning benefits but at a cost to engagement \cite{pardos2023learninggaindifferenceschatgpt, 10.1145/3698205.3733960}. Google DeepMind’s LearnLM team has produced a series of evaluations during the development of a tutorial chatbot \cite{jurenka2024responsibledevelopmentgenerativeai}and, in partnership with Eedi, conducted a small (N = 165) RCT reporting a 5.5\% improvement in independent problem solving over human tutoring alone (93.6\% probability of a genuine improvement) \cite{learnlmteam2025aitutoringsafelyeffectively}. However, these evaluations remain proprietary and focus narrowly on tutorial interaction, privileging instruction-following over holistic pedagogical adaptivity \cite{LearnLM2025GeminiArena}. 

Other efforts are noteworthy but limited. \citet{xu2025edubenchcomprehensivebenchmarkingdataset}’s general-purpose AI education evaluation lacks grounding in widely accepted pedagogical frameworks. Oak National Academy released a benchmark dataset for AI-generated educational content safety, though its scope is restricted to safety concerns \cite{clark2025autoevaluation}. \citet{Kennedy2026Vernacularized}’s taxonomy of AI harms in education advances context-specific evaluation but has yet to be widely operationalized. \citet{shetye2024khanmigo} qualitative analysis of Khanmigo using \citet{chapelle2001computer}’s CALL framework offers useful insights but relies on personal experience and thus remains anecdotal.

\section{Building L2-Bench}
\subsection{Theory and design}
A gap remains. Current evaluations insufficiently cover educational contexts, particularly second-language learning, and frequently misgeneralize tutorial interaction to group instruction. We present the first AI evaluation methodology for language learning to our knowledge. It is also among the first \textit{holistic} evaluations of AI in educational contexts that evaluates AI models at both the instance- and systemic- level \cite{10.1093/oxfordhb/9780198940272.013.0025}. 

Context-specificity is key. Language education is different from most other areas of education: it requires learners to acquire more implicit knowledge and proceduralised skills than other subjects; it comprises both the target of learning and the means of learning; it is heavily influenced by affective factors (motivation, identity, anxiety, confidence, and willingness to communicate) \cite{papi2023second}; and it is fundamentally shaped by each learner’s own social and cultural experience \cite{poehner2024sociocultural}. As a result, language is never “solved” \cite{dekeyser2025skill}.  

Furthermore, classroom learning is not the aggregation of individual interactions; knowledge is produced through social interaction \cite{bandura1977social, Kennedy2024Vernacularizing}. Because AI systems shape learning both directly and indirectly through instructional materials, evaluation must extend beyond subject-matter knowledge to capture learning experience design \cite{knight2026cecinestpasune, Kennedy2026Vernacularized}. Indeed, our evaluation is more interested in assessing LLM capabilities in designing experiences that promote the “doing,” not diagramming, of language \cite{Searle1996-SEAWIL, austin1975how}. 

As the contours of language learning are unique, we aim to produce evaluation artefacts that are each representative of the peculiar “vernacular” \cite{Kennedy2024Vernacularizing} of language learning. We draw upon three frameworks common to UK, EU, and global language learning design: the Council of Europe’s Common European Framework of Reference for Languages (CEFR) \cite{councilofeurope2001cefr}, the Eaquals Framework for Language Teacher Training and Development \cite{eaquals2016framework} and the British Council's Continuing Professional Development Framework \cite{BritishCouncil2025CPD}. We also engage experts in a series of practitioner validation exercises. 

The pedagogical frameworks utilized in this study correspond to the leading UK/EU frameworks for teaching these varieties of English. We utilize these frameworks here to demonstrate the importance of these components to the proposed methodology. In future implementations of this method, other global frameworks of importance could likewise be used to produce evaluation tasks representative to the target teaching context. Note that Eaquals reflects global L2 education best practices.
Lastly, we should restate that our benchmark initially intends to assess model performance specifically on EFL education (English as a Foreign Language), in the medium of US or UK varieties of English (see Appendix C.1 for more). We do so because these varieties of English represent the most-often taught second languages across the world \cite{blanco2025duolingo}. We recognize that models do not perform equally well across World English varieties \cite{smart2024sociallyresponsibledatalarge}. 

\subsection{Components}

\subsubsection{Taxonomy}

To evaluate AI capabilities in language education learning design, we first define a "learning experience designer in second language education" which encompasses the range of roles that intentionally design the conditions that shape how people learn: teachers, materials developers (content or assessment creators), learning designers, and teacher trainers (see Appendix C.1 for glossary). We then define the construct as a hierarchical competency taxonomy that articulates the capabilities required for effective “learning experience design” in L2 education. 

Our cross-disciplinary team initially developed the taxonomy under three constraints: (1) understandability – avoiding deep hierarchies that obscure interpretation; (2) independence – reducing ambiguity in task classification while acknowledging overlap; and (3) practitioner credibility – we draw upon several established pedagogical frameworks adapted interactions with AI systems. 

It is useful to distinguish between two complementary frameworks for language learning: a knowledge framework representing what learners need in order to learn (i.e. capturing the science of how language learners learn), and a competency framework representing the application of that knowledge (i.e. what practitioners need to do to apply knowledge effectively). They are critical distinctions: L2-Bench is based on the latter; we do not merely benchmark pedagogical knowledge as existing evaluations already show saturation on knowledge-based tasks \cite{lelièvre2025benchmarkingpedagogicalknowledgelarge}. 

The current L2-Bench taxonomy comprises 12 competencies and 31 sub-competencies organized into 12 main competencies across a two-level hierarchy (Figure 1) that span the full scope of a \textbf{second language learning experience designer in second language education} construct. 

\begin{table}[H]
\centering
\caption{L2-Bench taxonomy competencies.}
\label{tab:l2bench-taxonomy}
\begin{tabular}{ll}
\toprule
\textbf{Number} & \textbf{Competency} \\
\midrule
1  & Course Planning \\
2  & Lesson Planning \\
3  & Activity Planning \\
4  & Language Presentation \\
5  & Activity Management \\
6  & Exchange Partner \\
7  & Performance Evaluation \\
8  & Giving Feedback \\
9  & Progress Tracking \\
10 & Emotional Intelligence \\
11 & Assessment Creation \\
12 & Professional Development \\
\bottomrule
\end{tabular}
\end{table}

Each competency comprises one to six sub-competencies that capture specific capabilities assessable through “consensus criteria” (Components). For example, to be competent in "Giving Feedback" requires four sub-competencies: identifying errors and diagnosing their causes; prioritizing areas for feedback; providing explanations, models, or hints; and providing improvement activities. See Appendix C.2 for the full L2-Bench competency taxonomy.

\begin{figure}[h]
    \centering
    \includegraphics[width=.85
    \linewidth]{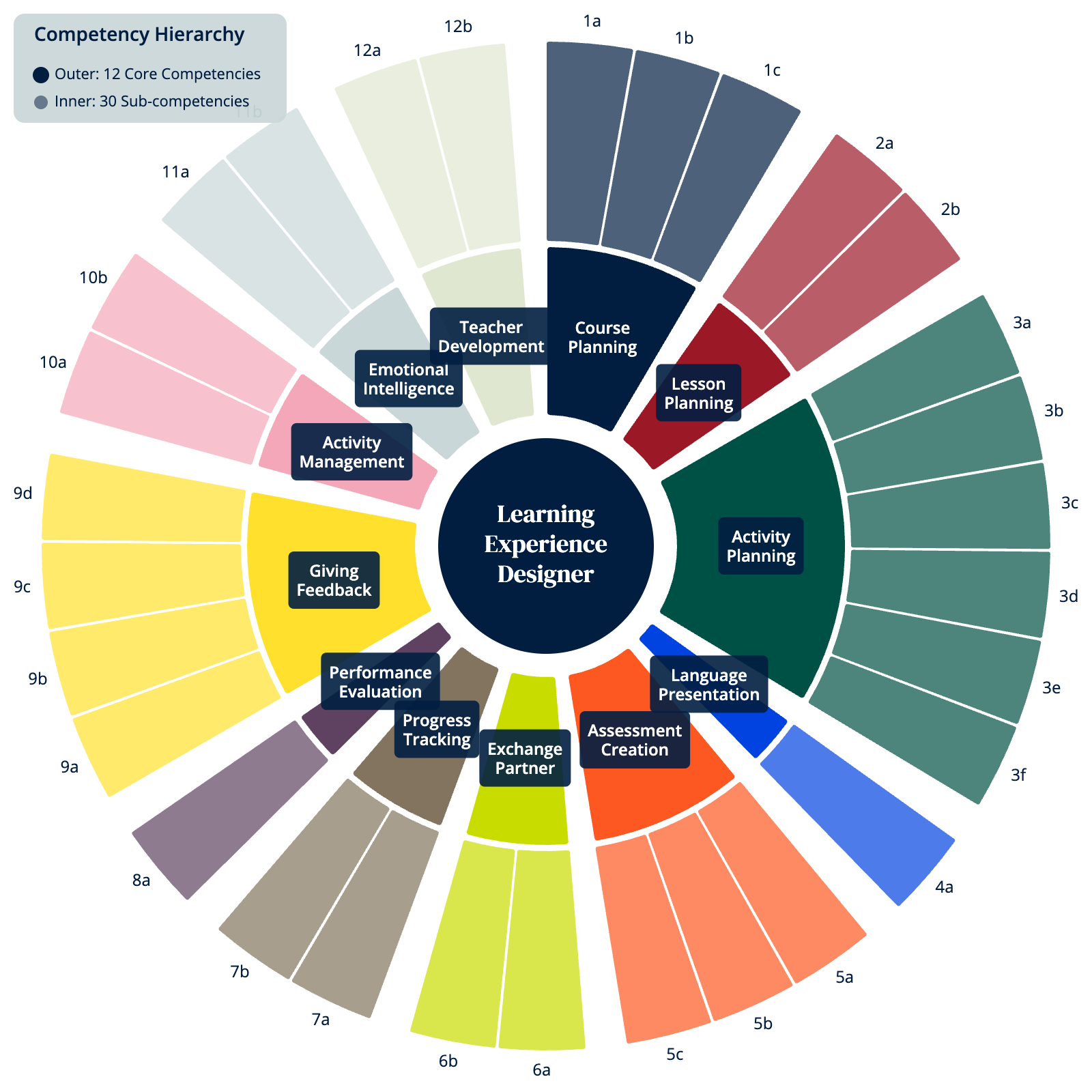}
    \caption{L2-Bench Taxonomy of Competencies - sunburst visualization showing the 12 competencies and 30 sub-competencies of a “learning experience designer in second language education.” See Appendix C.2 for large format figure.}
    \label{fig:taxonomy_sunburst}
\end{figure}

Validation of the taxonomy proceeded through multiple stages: iterative review during dataset construction of the tasks and criteria (Components); the pilot validation (Pilot Validation Exercise) providing empirical signal on whether tasks designed around the taxonomy measured coherent constructs; and forthcoming practitioner validation with representative stakeholder groups (Future Work). 

\subsubsection{Dataset}
The quality of an evaluation benchmark hinges on the questions it asks. To this end, we operationalize our competency taxonomy to design task-response pairs that assess capabilities against each competency, targeting over 1,000 tasks to create a high-quality dataset with meaningful variation. Task design follows four heuristics adapted from the UK AI Safety Institute's guidance on question-answer pair development \cite{aisi2024early_insights_qa} and recently proposed AICALL frameworks \cite{bahari2025call}: 

\begin{itemize}
    \item \textbf{Relevance}: Tasks are authentic real-world scenarios that directly relate to “learning experience designer competencies”, therefore requiring open responses.
    \item \textbf{Perspective}: Tasks consider multiple stakeholder viewpoints with full coverage of educational contexts.
    \item \textbf{Clarity}: Tasks include sufficient guidance to enable appropriate responses while avoiding ambiguity that could lead to inconsistent scoring.
    \item \textbf{Originality}: Tasks test application of pedagogical knowledge to new situations rather than relying on memory.
\end{itemize}

We constrained tasks to single-turn conversations (task-response pairs) to enable simplicity in data creation and evaluation. Multi-turn alternatives involve difficult experimental setup (recruiting users; contrived scenarios of experts role-playing as users; or unvalidated synthetic users) as well as requiring both turn-level and conversation-level evaluations that increase complexity. We acknowledge this constraint trades off against real-world relevance (learning interactions are inherently multi-turn), and we actively plan to expand into multi-turn dataset production. However, we note that many \textit{learning design} tasks are appropriately represented as single turn interaction (e.g. an instructor prompting an LLM for a quick introductory activity). And futhermore, beginning with the simplest possible form of interaction (adjacency pairs) provides a stable baseline, which not only aids iterative development but also project feasibility.

The dataset comprises tasks distributed across the 12 competencies that span diverse teaching contexts (“task variables”): geographic regions (Far East, South-East Asia, Middle East, Latin America, Europe, Africa), learner profiles (ages from 4–6 to 26+, CEFR levels A1–C2, primary school to corporate training settings), and learning aims (academic, professional, exam preparation, travel, cultural). 

To ensure scalability of generation and maintain experimental control, L2-Bench items are produced through a hybrid human-AI authoring approach that is modelled on publishing workflows: 

\begin{enumerate}
    \item “design”: language pedagogy experts create hand-crafted task exemplars and establish prompt templates for both task and reference answer creation
    \item “draft”: state-of-the-art foundation models with agent scaffolding generate candidate tasks, task criteria, and reference answers using the prompt templates and examples
    \item “review”: experts iteratively refine generated content, with modifications triggering regeneration cycles
    \item “approval”: a separate expert validates the reviewed items for pedagogical soundness
    \item “publish”: items are published in version-controlled benchmark dataset release (see Appendix D.1).
\end{enumerate}

To create authentic practitioner scenarios, some tasks include "resources" (documents in markdown or CSV), however since our goal is not to benchmark tool-handling phenomena, we simply append these resources in-context. Representative task examples are provided in Appendix D.2. 

Dataset validation proceeds through multiple stages: internal error analysis examines tasks for unrealistic scenarios, sociocultural biases, and scoring ambiguity; pilot data validation (Pilot Validation Exercise) provides quantitative signal on task authenticity and criteria quality for 325 task-response pairs, as well as qualitative reports that inform the next stage of iterative refinements; and future practitioner data validation (Future Work) following scaling to the full 1,000+ item dataset. 

\subsubsection{Measures}

Since tasks are designed to elicit open responses, evaluation cannot rely on standard accuracy metrics. We employ rubric-based measurement common to high-stakes, contextual domains (e.g. medical diagnosis) \cite{arora2025healthbenchevaluatinglargelanguage}.

\paragraph{Binary Criteria.} Scoring rubrics are composed of weighted criteria, where each criterion outlines what an ideal response should include or avoid. We employ binary pass/fail decisions for each criterion rather than Likert scales for two reasons: (1) calibrating both human annotators and LLM-Judges ("auto-scorers") on multi-point scales where the differences between adjacent points are subtle is challenging \cite{yan2025productevals}; (2) binary labels yield faster, more consistent human annotations, and simpler auto-scorer alignment due to clear decision boundaries. Criteria are assigned point values from -10 to +10 based on importance, with negative points for undesirable responses. Final scores are computed by summing points for criteria met and dividing by the maximum possible score.

\paragraph{Criteria System.} The hierarchical structure of the competency framework enables granular assessment while maintaining connection to broader pedagogical competencies. We define three types of criteria:

\paragraph{Consensus Criteria.} Defined per sub-competency with expert agreement, capturing essential requirements for competent performance. Tasks inherit consensus criteria from their tagged sub-competencies. For example, any task tagged with sub-competency 08a (identifying errors and diagnosing causes) includes the consensus criterion: "Estimates the likely causes of the error - e.g. gaps in knowledge, or skill proficiency" (weight: +5).

\paragraph{Task Criteria.} Criteria specific to individual tasks that capture context-specific requirements (designed to be unique).

\paragraph{Universal Criteria.} Criteria applied to all tasks, with weightings conditional on task context, for example, CEFR-level language appropriateness (weight: +9 for learner-facing responses, +2 for teacher-only responses) and child-safety considerations (weight: -10 for offensive content, -5 for sensitive content requiring teacher guidance).
\newline

A task rubric is therefore composed of task criteria, consensus criteria, and universal criteria, all designed to be independent of one another (see Appendix C.2 and Appendix C.3). Reference answers are designed to score maximally against each task rubric, providing guidance for both human and automated scoring.

\subsubsection{Scoring pipeline}

We do not want to underestimate what models can do, but instead elicit their best response, and therefore every task has a system prompt to capture the implicit context a practitioner with expertise in the underlying competency may have to perform the task, without revealing task scoring rubrics. We design system prompts using best practices in capability elicitation \cite{ukaisi2024elicitation} pairing each task with its own unique prompt template of: 

\begin{enumerate}
    \item role establishment
    \item domain expertise
    \item contextual framing (based on the “task variables”)
    \item optional chain-of-thought reasoning.
\end{enumerate}

We recognise that optimal elicitation techniques vary across models, but we accept this trade-off vs unelicited prompts that may underestimate capability or even favor certain models over others due to inherent prompt biases \cite{abbas2025developingmaintainingopensourcerepository}.

We use open or cloud-hosted API endpoint models for all generation and scoring pipelines in order to minimise leakage to model providers and therefore mitigate benchmark saturation and contamination. A model receives the task-specific system prompt and the task itself (with any "resources") as input and generates an open-response.

Scoring open-responses against rubrics at scale requires automated approaches; we therefore employ LLM-as-a-Judge (auto-scorer) as our scoring mechanism for assessing open-ended generations against our task rubrics. Having deconstructed our rubrics into binary decisions, we can present simpler judgments to the auto-scorer by providing it with a single criterion at a time \cite{yan2025productevals}, whilst also mitigating (but not solving) documented limitations in LLM-as-a-Judge literature, such as verbosity bias (preferring longer responses) and judge bias (favouring responses from the same model family) \cite{ukaisi2025llm}. We will employ techniques to test for known failure modes that our research design implicates most clearly, such as pairwise testing \cite{liu2024aligning} and playing favorites \cite{spiliopoulou2025playfavoritesstatisticalmethod}.

We develop our auto-scorer in two phases: (1) initially, we use Claude Sonnet-4.5 with thinking as the auto-scorer foundation model for our scoring pipeline (outputting reasoning traces to enable subsequent meta-analysis), then (2) we will later collect scores from human experts in our future practitioner data validation and use this data to optimise the auto-scorer by varying models and prompts.

Currently, we employ reference-guided prompting for our auto-scorer, whereby the auto-scorer receives the task-response pair along with the original task context, task metadata (competency, sub-competency, task variables), the reference "gold standard" answer, and a single criterion from the task rubric to score against. This approach assumes that the reference answer is an objective solution. Of course, this does not hold in many educational contexts, and so we temper our expectations for inter-judge agreement in our future practitioner data validation (Future Work).

Our scoring pipeline begins with generating a response per task. For each response, our auto-scorer determines whether the response meets each criterion; if the criterion is met, full points are given, otherwise no points are given. We then get the total points for a given task-response pair by summing the point values for criteria met and dividing by the maximum possible score to produce the task score (a task score can be negative if more negative points were assigned to it than positive points). For each task, we calculate an AI system's overall L2-Bench score by taking the mean of these task-level scores, clipped between 0–100

To ensure trust in our future L2-Bench leaderboard rankings, we will conduct uncertainty quantification that distinguishes genuine performance gaps from measurement noise \cite{miller2024addingerrorbarsevals}. We will therefore generate n = 3 responses per task and compute the mean score, reporting standard errors alongside point estimates to enable statistical inference on model differences rather than relying solely on rank ordering (see Appendix E for details). 

\section{Pilot Validation Exercise}
Prior to full-scale development of L2-Bench, we conducted a pilot validation study (IRB-exempt) which served two purposes: 
\begin{enumerate}
    \item \textbf{gathering early feedback} to improve the competency construct and dataset design before scaling to the full 1,000+ tasks
    \item \textbf{informing best practices} for a future global practitioner data validation (Future Work), such as establishing evaluation design parameters, identifying challenging competencies, and calibration guidelines.
\end{enumerate}

\subsection{Methods}

Participants (N = 39) were recruited from a leading UK university in collaboration with the university’s careers network team. Recruitment proceeded via an intranet announcement for voluntary work experience that included an application form detailing: eligibility (postgraduates enrolled in taught masters programmes in Arts, Humanities and Social Sciences), time commitments (2-3 hours per week over six weeks) and the study format (a team-based ‘challenge’). 

While this participant group was chosen primarily for availability within our development timelines, their critical thinking backgrounds and diverse perspectives were deemed appropriate to scrutinise our benchmark components to identify patterns that would allow us to iterate. Furthermore, despite not holding practitioner roles that would be best placed to validate L2-Bench (see Section 5), 32 of the participants were international students with L2-English proficiency, 8 had prior teaching experience, and 4 were enrolled on MA Education programmes, all relevant experiences direct or adjacent to L2 education that could be reasonably drawn upon for initial validation signal on elements of our benchmark. 

Participants were initially put into 8 teams of 6 stratified by L2-English and teaching experience to balance expertise levels (see Appendix A.2 for complete team composition), with the challenge format (involving prize incentives) encouraging teams to operate independently as cluster-level evaluators (N = 8), minimising leakage between samples. Teams were assigned 325 task-response pairs in randomised order across each team to ensure broad coverage across all competencies. Data (complete with all task metadata, reference answer) was provided in both Excel format and on an internally-hosted instance of the open-source Langfuse annotation platform (which was also used to collect item-level ratings and free-text annotations) to accommodate different working preferences. Task responses were generated with Claude Sonnet-3.7 via Amazon Bedrock API with default settings (T = 1.0, top-p = 0.999, top-k = 250). 

To measure dataset validity, we encouraged participants to rate task authenticity and criteria adequacy for each task item in the dataset on 5-point Likert scales (we used rather than 7-point scales to enable faster ratings without sacrificing reliability \cite{preston2000optimal}): 

\begin{itemize}
    \item \textbf{Authenticity score}: Does the task represent an authentic L2 educational scenario?
    \item \textbf{Criteria score}: Do the task rubric criteria evaluate the AI response well enough to distinguish good vs poor responses?
\end{itemize}

Beyond individual ratings, teams produced written reports identifying systematic issues in task design and broader competency construct and proposing revisions. Teams received five hours of dedicated in-person training, with two 2-hour sessions dedicated to study orientation and calibration (objectives, background context, interface, practice task validation examples, norming discussions) within the first two weeks, plus one additional hour per team for focused question-and-answer support in the third week. 

\subsection{Results}
\subsubsection{Response rates and coverage}

The study collected 1,128 ratings across 325 unique tasks, achieving an overall response rate of 43\% (1,128 of 2,632 possible rating opportunities). Response rates varied considerably across teams, ranging from 15\% to 89\% (see Appendix A.3), reflecting a combination of team  dynamics, the substantial time required per task (teams estimated 10–20 minutes for the average task) and poor UI experience on the annotation platform (teams preferred the spreadsheet interface). Critically, there was 100\% task coverage (all tasks received at least one rating), with 77\% of tasks receiving three or more independent ratings - the minimum typically required for meaningful reliability analysis. Response patterns varied systematically across competency domains, revealing which areas evaluators found more accessible (see Appendix A.4). Certain responses were missing but not at random, with specialised competencies showing notably higher skip rates: "Evaluate student's performance" (69\%) and "Support professional development" (63\%). Given teams had varying levels of experience with AI tools and pedagogical assessment, this self-selection pattern may indicate that only evaluators feeling qualified responded, potentially concentrating expertise in those ratings.

\subsubsection{Dataset validity}
After converting the 5-point Likert scales (where 1='Strongly Disagree' through 5='Strongly Agree'), evaluators broadly endorsed benchmark tasks as representing realistic teaching scenarios, with authenticity scores averaging mean M = 4.24 (95\% CI [4.19, 4.30]). This offers some evidence that our task design methodology reflects authentic teaching practice, albeit the UK-based university evaluators limits generalisability to global L2-educational contexts. However, criteria scores were lower (M = 3.93, 95\% CI [3.87, 3.98]). The Wilcoxon signed-rank test confirmed this gap (W = 42,690, p $<$ 0.001, d = 0.28), indicating that while participants agreed that the tasks are realistic, they are less sure that the criteria are sufficient to evaluate pedagogical quality of AI responses. Table 2 gives the statistical summary of the pilot data validation (see Appendix A.1 for complete statistical methodology).

Performance varied meaningfully across competency domains, with pedagogically complex competencies such as "Present language learning points" (M = 3.69) and "Giving feedback" (M = 3.77) scoring lowest on criteria scores despite high authenticity ratings (see Figure 2). 


\begin{table}[h]
\centering
\caption{Summary statistics from pilot data validation. Auth=Authenticity; Crit=Criteria; M=Mean; CI=95\% confidence interval; SD=Standard deviation; IAA=Inter-annotator agreement (Krippendorff's $\alpha$) for critera; IIC=Internal item consistency (Cronbach's $\alpha$) for criteria.}
\label{tab:1-pilot-summary}
\begin{tabular}{lc}
\toprule
\textbf{Metric} & \textbf{Value} \\
\midrule
Tasks & 325 \\
Ratings & 1,128 \\
Skip \% & 57\% \\
Auth M & 4.24 \\
Auth CI & [4.19, 4.30] \\
Crit M & 3.93 \\
Crit CI & [3.87, 3.98] \\
Crit SD & 0.99 \\
IAA ($\alpha$) & $-$0.01 \\
IIC ($\alpha$) & 0.95 \\
\bottomrule
\end{tabular}
\end{table}

\begin{figure}[h]
    \centering
    \includegraphics[width=1.0\linewidth]{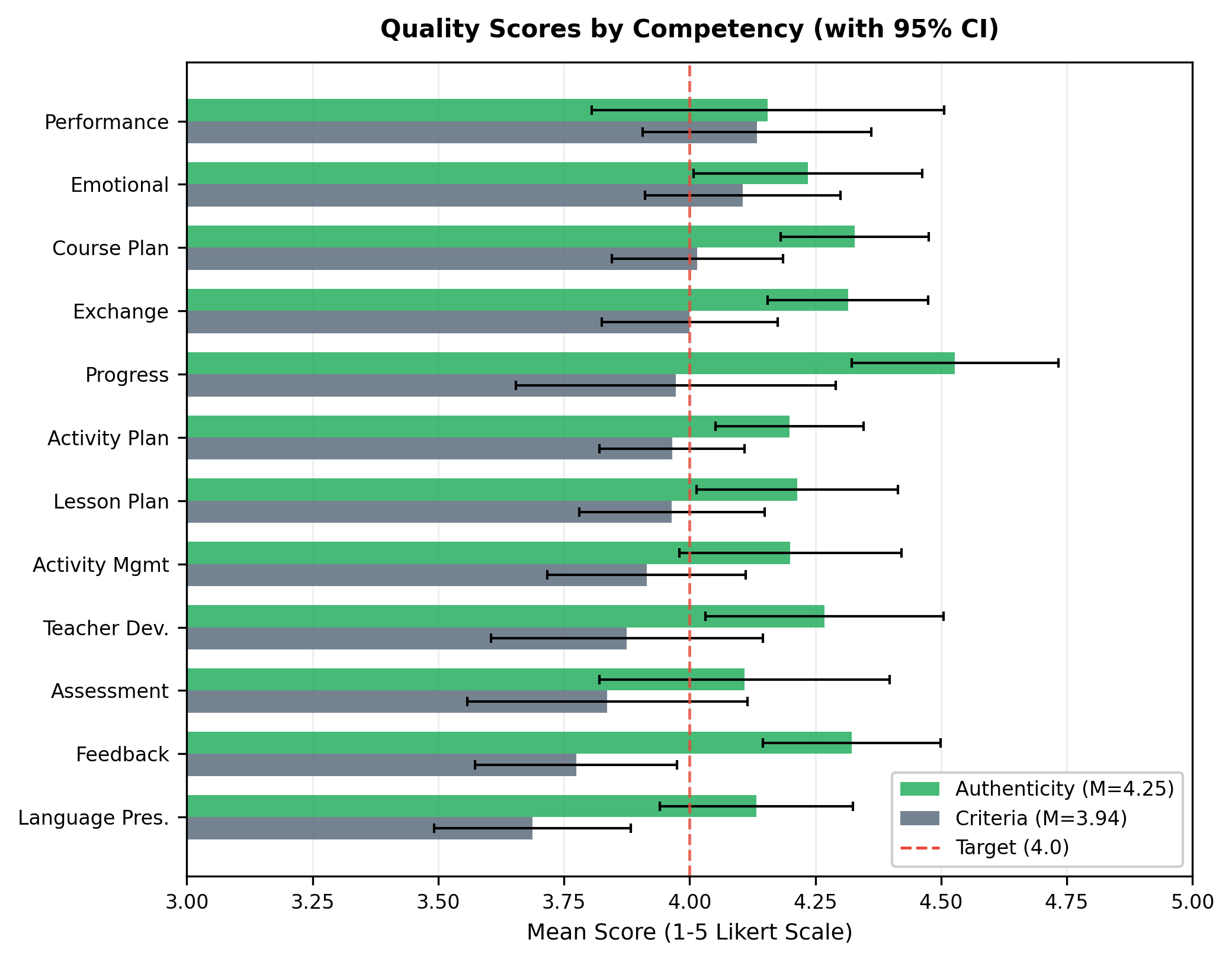}
    \caption{Criteria and authenticity scores per competency with 95\% CIs.}
    \label{fig:placeholder}
\end{figure}

  \begin{figure}[h]
    \centering
    \includegraphics[width=1.0\linewidth]{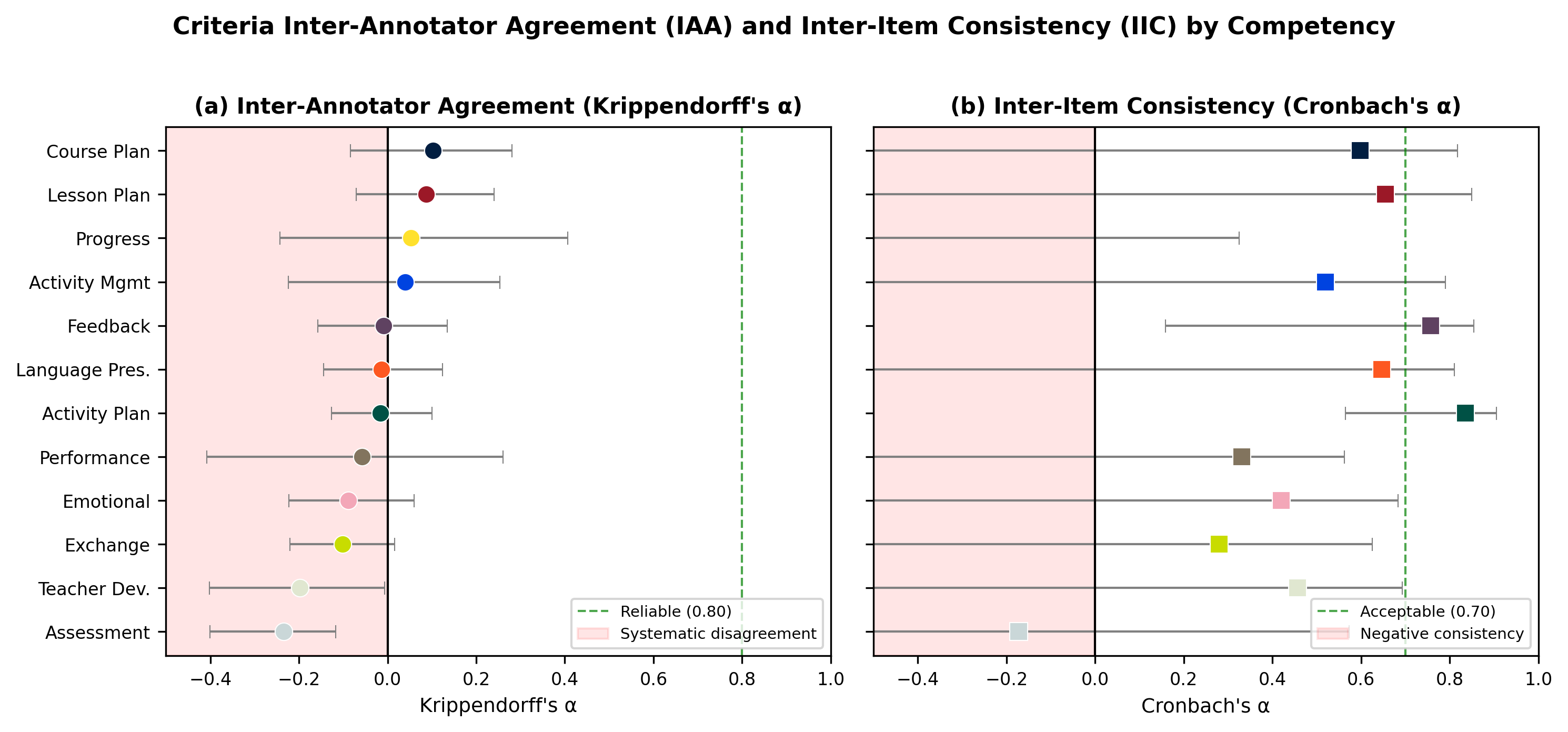}
    \caption{Side-by-side panels showing (a) Krippendorff's $\alpha$ (criterion IAA) and (b) Cronbach's $\alpha$ (criterion IIC) by competency for criteria scores with 95\% CI.}
    \label{fig:placeholder}
\end{figure}

We measured inter-annotator agreement on criteria (hereafter criteria IAA) using Krippendorff's Alpha to handle incomplete data (see Appendix A.1). Criteria IAA scores were universally poor across all competencies (see Figure 3). Two-thirds (8/12) showed negative values, indicating systematic disagreement where evaluators diverged more than expected by chance. The maximum alpha achieved was just 0.10 (Course Planning), far below the 0.80 threshold required for reliable conclusions  \cite{krippendorff2004content}. 

We also conducted criteria internal item consistency (hereafter criteria IIC) analysis using Cronbach's Alpha to assess whether tasks within each competency measured a coherent underlying construct (see Appendix A.1 for details). Most competencies demonstrated at least low-to-moderate criteria IIC values ($\alpha$ $\geq$ 0.40) on criteria scores, with overall criteria IIC achieving excellent reliability ($\alpha$ = 0.95) (see Figure 2). 

This pattern of poor criteria IAA with low-to-excellent criteria IIC is consequential. It provides initial evidence that tasks designed around the competency taxonomy are \textit{measuring the same underlying concept}, even if evaluators \textit{apply systematically different standards} to assess the suitability of proposed ways of measuring that concept. 

We can explain this result primarily via calibration: ultimately evaluators were postgraduate students, not experienced practitioners, meaning that evaluators brought their own implicit standards for "good" performance. Future validation exercises with professional practitioners will shed more light on this phenomenon (Future Work). At the same time, we felt it important to report our negative results against criteria to demonstrate to those working in allied fields that they should exercise caution when transferring rubric-based evaluations methodologies across domains. Ultimately, we think this also factors into the misalignment between poor criteria IAA on highly-rated criteria IIC items.

The "Giving Feedback" competency warrants mention as it illustrates the distinction between construct validity problems and evaluator calibration problems. This competency showed the highest score variance (SD = 1.13) and the second-lowest mean criteria score (M = 3.77) despite high authenticity (M = 4.32). Yet it exhibited the second-highest Cronbach's alpha ($\alpha$ = 0.76) alongside negative Krippendorff's alpha ($\alpha$ = -0.01), following the overall trend that tasks within "Giving Feedback" measure a coherent construct, but that evaluators apply systematically different standards when judging feedback criteria.

Nevertheless, despite poor criteria IAA scores, moderate to excellent criteria IIC scores from an adequate sample of one of our core stakeholder groups (postgraduate learners) provide \textbf{promising early signal that the L2-Bench construct is sound}. We propose future and larger-scale validation work to confirm these findings (Future Work).

\subsection{Iteration}
Beyond quantitative analysis, teams produced written reports identifying systematic issues across L2-Bench components. Prior to the scaling up to the full 1,000+ task dataset, we may use this feedback to make revisions in three areas: 

\paragraph{Competency taxonomy.} Several sub-competencies were identified as oversimplified. In other areas, practitioners observed “inter-competency” elements: capabilities that appear important in multiple competencies. We will determine whether these represent distinct competencies or cross-cutting capabilities requiring separate treatment. We hypothesize that the method (of producing a taxonomy based on language-specific pedagogies) will scale to other pedagogies. However, we also believe each of these is, ultimately, an empirical question and one we will consider in future work.

\paragraph{Evaluation criteria.} Evaluators noted that some criteria are more abstract or more context-sensitive than others; we plan to implement variable consensus criteria weightings based on task context and expand universal criteria to address sensitivities that vary by geographic and cultural context. 

\paragraph{Task design.} We will expand on our task variable approach for systematic task design, creating a framework of the context dimensions that impact on language learning tasks and responses – from linguistic context, to educational, resource availability and social factors. Additionally, we will standardise task wording for consistency where equivalent.

\section{Future Work}
The tasks of a "learning experience designer in second language education" span multiple roles (teaching, content creation, assessment, learning design, professional development), but no single person or role is sufficiently skilled in all areas to validate these independently. Even if such a person existed, the dynamics of global pedagogy vary significantly from person to person, day to day, learning problem to learning problem. Therefore, we need data validation that brings together multiple stakeholder groups: 

\begin{itemize}
    \item \textbf{Language learning practitioners}: Experienced practitioners that design the conditions in which people learn
    \item \textbf{Language teachers}: L2 teachers, particularly those who engage with pedagogical position papers and research
    \item \textbf{Language assessment specialists}: Professionals who design and validate L2 assessment tests
    \item \textbf{Researchers}: Academic researchers in L2 education
    \item \textbf{EdTech professionals}: Professionals who design, implement or administrate  applications for L2 education
    \item \textbf{Learners}: Advanced adult L2 learners for language fluency requirements and ethics considerations.
\end{itemize}

To this end, we will build on our pilot data validation to conduct a global "practitioner data validation" study, recruiting volunteers across the above stakeholder groups from our institution network whilst accounting for practical constraints in both representation (see Appendix B2) and ethical considerations (see Impact Statement). Practitioners will be assigned to "practitioner groups" (groups) matching their expertise through a short pre-screening survey, and these groups are then mapped to primary competencies from the L2-Bench competency construct which will be used to filter which tasks they will validate (see Appendix B2 Table 8 for group definitions and competency coverage). This stratified approach ensures specialist coverage for challenging competencies, while enabling cross-validation by practitioners with diverse professional perspectives. 


\begin{table*}[ht]
\centering
\caption{Research objectives (RO), statistical methods, success criteria and power status (\checkmark = well-powered at 80\%) for practitioner validation. IAA and IIC computed separately for authenticity and criteria scores.}
\label{tab:2-research-objectives}
\footnotesize
\begin{tabular}{llllcc}
\toprule
RO & Analysis & Methods & Target & In Pilot & Power \\
\midrule
RO1 & Authenticity M & One-sample $t$-test & $>$4.0/5 & \checkmark & \checkmark \\
RO1 & Criteria M & One-sample $t$-test & $>$3.5/5 & \checkmark & \checkmark \\
RO1 & IAA & Fleiss' $\kappa$ & $>$0.20 & Kripp's $\alpha$ & \checkmark \\
RO1 & IIC & Cronbach's $\alpha$ & $>$0.70 & \checkmark & \checkmark \\
RO1 & Auth vs Criteria gap & Wilcoxon signed-rank & --- & \checkmark & \checkmark \\
RO1 & Mixed effects model & Likelihood ratio test & $d$=0.30, $p<$0.05 & --- & $\sim$45\% \\
RO2 & A/B preference & Binomial test & $\sim$70\% ref, $p<$0.05 & --- & \checkmark \\
RO3 & Inter-judge agreement & Cohen's $\kappa$ & $>$0.60 & --- & \checkmark \\
RO3 & Auto-scorer sensitivity & Recall & $\geq$0.80 & --- & \checkmark \\
RO4 & Group differences & Mann-Whitney $U$ & $r$=0.3 & --- & $\sim$68--92\% \\
\bottomrule
\end{tabular}
\end{table*}

The practitioner data validation will involve the full L2-Bench dataset of 1,300 task-response pairs (1,000 pairs to be released, 250 hold-out pairs for auto-scorer optimisation and benchmark saturation detection, and allowing for up to 50 pairs that may be excluded following validation (see Appendix B.5)), addressing three primary research objectives (ROs): 

\paragraph{RO1. Measuring dataset validity} by establishing practitioner agreement on task authenticity and criteria adequacy.

\paragraph{RO2. Measuring answer quality} by conducting blind comparisons of AI-generated responses to tasks versus our reference answers.

\paragraph{RO3. Measuring auto-scorer validity} by collecting practitioner scores on “AI answers” against the task rubric and establishing inter-rater agreement.
\newline

We will also explore whether agreement varies systematically by practitioner expertise on competencies where we have diverse professional coverage (RO4). 

While RO1 mirrors our approach to the pilot data validation (Pilot Validation Exercise), RO2 and RO3 are motivated by our desires to measure the reference answers produced in our task item production process (Components) and to explore the extent of practitioner bias against AI answers over reference answers. Moreover, the data collected for RO3 will be used to optimize our auto-scorer before dataset release. 

For each task-response pair shown to the practitioner, the validation workflow proceeds sequentially through 2-3 stages: 

\begin{enumerate}
    \item practitioners rate task authenticity and criteria adequacy on 5-point Likert scales
    \item they complete blind A/B comparisons where two answers – one AI-generated, one reference – are presented in randomized order; practitioners indicate preference with optional comments, and
    \item after revealing the "AI answer" (which is randomized to be either the AI response or our reference answer), some validators score the "AI answer" against the task rubric on each criterion as Pass/Fail with optional comments, enabling measurement of inter-rater agreement with the auto-scorer.
\end{enumerate}

The AI response will be generated by a single frontier model using the model provider's recommended default settings. 

To achieve well-powered analysis, we seek 5 validity raters per task (Part 1+2) and 3 judge raters per task (Part 3). We analyze results using Fleiss' Kappa for inter-annotator agreement (IAA), Cronbach's Alpha for internal consistency (IIC), Cohen's Kappa for inter-judge agreement (IJA), and mixed-effects models to test whether ratings differ systematically across competencies while controlling for rater and item effects. Table 2 summarizes our research objectives, statistical methods, and success criteria (full statistical methods and power analysis are detailed in Appendix B.3).

Based on pilot results, we anticipate authenticity ratings exceeding M = 4.0/5 and criteria adequacy exceeding M = 3.5/5, strong IIC $>$ 0.70, and variable Fleiss IAA $>$ 0.20 across competencies - pedagogical quality is  contingent upon many factors, including teaching philosophy and experience, learner needs, and the learning problem in question.  

We hypothesize reference answers will be systematically preferred over AI answers when that is true ($\sim$70\% preference for reference), but when the AI answer is disguised as a Reference answer, we suspect a smaller systematic preference for the "Reference answer", accounting for practitioner bias against AI answers. Furthermore, we set a pragmatic target of inter-judge agreement (IJA) for fair agreement ($\kappa$ $>$ 0.20), acknowledging the limitations of our initial reference-guided auto-scorer, recognising that many scenarios in language learning lack clear "correct" answers. For instance, when giving feedback, multiple lexical or semantic formulations may be acceptable, counterfactual learning techniques might be employed, and appropriacy will depend on unstated contextual factors \cite{knight2026cecinestpasune}, as well as accounting for practitioner fatigue \cite{yan2025productevals}. 

Given the substantial practitioner time commitment required to power our study ($\sim$ 2,400 hours), we document risk mitigation strategies in Appendix B.4, including over-recruitment buffers and reduced-coverage fallback designs that preserve statistical validity. In addition to validating L2-Bench, we note that we also use the results from the practitioner study to apply a task-response pair exclusion protocol (Appendix B.5) to remove validation outliers before the final dataset release \cite{truong2025fantasticbugsaibenchmarks}.

\section{Conclusion}
We have introduced L2-Bench, an evaluation benchmark intended to assess AI system capabilities for performing tasks entailed in quality learning experience design in second language education. Our paper primarily reported our methodology, detailed key artefacts (e.g. our taxonomy, measures, and aspects of our pilot dataset) and validation exercises in the hopes that these methods can be of broader use to the evaluations community as well as those developing AI systems for educational contexts, including but not limited to language learning. Ultimately, our work  demonstrates the feasibility of creating rigorous, scalable evaluations that bridge AI capabilities with learning science theory, contributing to efforts within the AI evaluation community to move beyond narrow accuracy metrics toward more rigorous, context-sensitive assessment of AI capabilities.

\section*{Acknowledgements}

We would like to thank all participants of the University of Birmingham ShapeAI Challenge for their hard work and valuable contributions to AI research. We would like to particularly thank the winners and runner-ups of the challenge: Venkata Vyjayanthi Pedapati (Vy), Yernur Niyetkaliyev, Aparajitha Magnesh, Manh Nguyen (Leo), Niamh Evans, Hsin-Yun Ho (Sydney), Sofía Muñoz, Saniya Saheer, Taiki Shimosakai, Yang Yu. 

We would also like to thank Dr. Liam Knight for his help in bringing the challenge from idea to reality, and for his tireless support in facilitating the challenge and assisting all participants for over 6 weeks. 



\section*{Impact Statement}

\subsection*{IS1. Human Subjects Research for Data Validation}

Both the pilot validation study and the forthcoming practitioner validation were designed in accordance with established research ethics principles. Participation is voluntary with explicit right to withdraw at any stage without penalty. Informed consent covers study purpose, procedures, time commitment, data handling, and anonymisation. All data collection complies with UK GDPR requirements; ratings are anonymised before analysis and stored securely following institutional data governance policies with 5-year post-publication retention. A formal legal agreement between participating institutions and/or individuals establishes clear provisions for intellectual property, confidentiality (5-year non-disclosure with specific AI tool restrictions to prevent data contamination that could compromise benchmark validity), and data protection.  

Both studies are structured as voluntary programmes. The pilot study was conducted in partnership with a UK university's careers network team as a voluntary challenge programme designed to provide meaningful professional development with non-monetary incentives, and a prize incentive for the winning team.The practitioner study will be conducted predominantly on a voluntary research basis with non-monetary incentives. All incentives are subject to compliance review to ensure fair recognition for participant contributions. 

The practitioner study additionally employs a blind A/B comparison protocol where participants compare two answers – one AI-generated, one reference – to whicih we then reveal the "AI answer" (which is randomized to be either the AI response or the reference answer). This methodologically necessary blinding is mitigated through full debriefing after completion, scientific rationale for the design (Future Work), and the opportunity to withdraw data post-debrief. We also address practitioner burden through flexible scheduling and clear time-per-task communication.

\subsection*{IS2. Broader Impacts}

We anticipate this work will contribute positively to the AI evaluation ecosystem by providing an open-source benchmark and methodology that enables more rigorous assessment of AI capabilities in educational contexts.

We remain attentive to several concerns. First, benchmarks assessing AI capabilities in education could, in principle, be repurposed to evaluate human educators; we note explicitly that our work focuses solely on AI system assessment and we have no products or interests in teacher evaluation. Second, our reliance on predominantly European frameworks (CEFR, Eaquals) may embed cultural assumptions that limit generalisability; we plan to partner with regional language education organisations to address this in future iterations. Third, leaderboard rankings could inadvertently incentivise benchmark-specific optimisation rather than genuine pedagogical improvement.

Unanticipated consequences may arise from applications of our competency taxonomy or evaluation methodology in ways we have not foreseen. We encourage researchers building on this work to consider the potential for dual-use applications, to examine their own assumptions about pedagogical quality, and to implement appropriate safeguards when deploying evaluation frameworks in educational contexts. Likewise, we are actively exploring how our benchmark can evolve to account for multi-turn nature of authentic pedagogical relationships.
 
Once we are satisfied with the effectiveness of L2-Bench as described here, we plan to expand the benchmark itself. Planned expansion includes (1) increasing the number of tasks to maintain task representativeness and depth and establishing regular update intervals; (2) exploring how best to include support for multi-turn interactions; (3) expanding coverage to image and audio modalities (which are common in language learning design); and (4) including more globally diverse practitioner communities and language education frameworks. We intend to achieve this latter goal primarily through leveraging extensive institutional relationships with communities of practice (broadly construed) across the world.
 
We also note that L2-Bench is the first evaluation benchmark in a planned AI-for-education evaluation ecosystem. We dwell on our evaluation methodology because we intend to scale our methods to support evaluation in other domains for which usage data indicates evaluation needs are particularly acute—for instance, in legal pedagogy.

\subsection*{IS3. Generative AI Statement}

Generative AI tools were used in several aspects of this research.

For core methodology, Claude models (Anthropic) via Amazon Bedrock API served three functions: (1) Claude Sonnet-4.1 and Sonnet-4.5 with Claude Code initially generated candidate tasks, criteria, and reference answers during hybrid human-AI dataset authoring (Components); (2) Claude Sonnet-3.7 generated AI responses for pilot validation (Methods); and (3) Claude Sonnet-4.5 with extended thinking serves as the auto-scorer foundation model (Components). All AI-generated content underwent expert review and validation.

For research support, GenAI assisted with: reviewing experimental design and identifying methodological improvements; research on statistical methods and their implementation; reviewing data processing pipelines and analysis iterations; and iterating on data visualisations.

For manuscript preparation, GenAI assisted with: LaTeX table and equation formatting, grammar and spelling review, and website development that resulted in figure creation.

All substantive research decisions, interpretations, and conclusions remain solely the responsibility of the authors. 

\bibliography{aaai2026}

@String{Computing = "Computing" }

@String{Computer = "{IEEE} Computer" }

@String{Chelsea = "Chelsea" }

@misc{xu2025edubenchcomprehensivebenchmarkingdataset,
  title={EduBench: A Comprehensive Benchmarking Dataset for Evaluating Large Language Models in Diverse Educational Scenarios},
  author={Bin Xu and Yu Bai and Huashan Sun and Yiguan Lin and Siming Liu and Xinyue Liang and Yaolin Li and Yang Gao and Heyan Huang},
  year={2025},
  eprint={2505.16160},
  archivePrefix={arXiv},
  primaryClass={cs.CL},
  url={https://arxiv.org/abs/2505.16160}
}

@article{clark-etal-2020-tydi,
    title = "{T}y{D}i {QA}: A Benchmark for Information-Seeking Question Answering in Typologically Diverse Languages",
    author = "Clark, Jonathan H.  and
      Choi, Eunsol  and
      Collins, Michael  and
      Garrette, Dan  and
      Kwiatkowski, Tom  and
      Nikolaev, Vitaly  and
      Palomaki, Jennimaria",
    editor = "Johnson, Mark  and
      Roark, Brian  and
      Nenkova, Ani",
    journal = "Transactions of the Association for Computational Linguistics",
    volume = "8",
    year = "2020",
    address = "Cambridge, MA",
    publisher = "MIT Press",
    url = "https://aclanthology.org/2020.tacl-1.30/",
    doi = "10.1162/tacl_a_00317",
    pages = "454--470"
}

@misc{miller2024addingerrorbarsevals,
  title={Adding Error Bars to Evals: A Statistical Approach to Language Model Evaluations},
  author={Evan Miller},
  year={2024},
  eprint={2411.00640},
  archivePrefix={arXiv},
  primaryClass={stat.AP},
  url={https://arxiv.org/abs/2411.00640}
}

@misc{aisi2024early_insights_qa,
  author       = {{UK AI Security Institute}},
  title        = {Early Insights from Developing Question-Answer Evaluations for Frontier AI},
  year         = {2024},
  month        = sep,
  day          = {23},
  url          = {https://www.aisi.gov.uk/blog/early-insights-from-developing-question-answer-evaluations-for-frontier-ai},
  note         = {Accessed: 2026-03-17},
  howpublished = {\url{https://www.aisi.gov.uk/blog/early-insights-from-developing-question-answer-evaluations-for-frontier-ai}}
}

@techreport{ukaisi2024elicitation,
  title={Elicitation of AI responses protocol},
  author={{UK AI Safety Institute}},
  year={2024},
  month={July},
  institution={UK AI Safety Institute},
  url={https://www.aisi.gov.uk/work/elicitation-protocol}
}

@misc{arora2025healthbenchevaluatinglargelanguage,
  title={HealthBench: Evaluating Large Language Models Towards Improved Human Health},
  author={Rahul K. Arora and Jason Wei and Rebecca Soskin Hicks and Preston Bowman and Joaquin Quiñonero-Candela and Foivos Tsimpourlas and Michael Sharman and Meghan Shah and Andrea Vallone and Alex Beutel and Johannes Heidecke and Karan Singhal},
  year={2025},
  eprint={2505.08775},
  archivePrefix={arXiv},
  primaryClass={cs.CL},
  url={https://arxiv.org/abs/2505.08775}
}

@misc{abbas2025developingmaintainingopensourcerepository,
  title={Developing and Maintaining an Open-Source Repository of AI Evaluations: Challenges and Insights},
  author={Alexandra Abbas and Celia Waggoner and Justin Olive},
  year={2025},
  eprint={2507.06893},
  archivePrefix={arXiv},
  primaryClass={cs.CL},
  url={https://arxiv.org/abs/2507.06893}
}

@misc{jurenka2024responsibledevelopmentgenerativeai,
  title={Towards Responsible Development of Generative AI for Education: An Evaluation-Driven Approach},
  author={Irina Jurenka and Markus Kunesch and Kevin R. McKee and Daniel Gillick and Shaojian Zhu and Sara Wiltberger and Shubham Milind Phal and Katherine Hermann and Daniel Kasenberg and Avishkar Bhoopchand and Ankit Anand and Miruna Pîslar and Stephanie Chan and Lisa Wang and Jennifer She and Parsa Mahmoudieh and Aliya Rysbek and Wei-Jen Ko and Andrea Huber and Brett Wiltshire and Gal Elidan and Roni Rabin and Jasmin Rubinovitz and Amit Pitaru and Mac McAllister and Julia Wilkowski and David Choi and Roee Engelberg and Lidan Hackmon and Adva Levin and Rachel Griffin and Michael Sears and Filip Bar and Mia Mesar and Mana Jabbour and Arslan Chaudhry and James Cohan and Sridhar Thiagarajan and Nir Levine and Ben Brown and Dilan Gorur and Svetlana Grant and Rachel Hashimshoni and Laura Weidinger and Jieru Hu and Dawn Chen and Kuba Dolecki and Canfer Akbulut and Maxwell Bileschi and Laura Culp and Wen-Xin Dong and Nahema Marchal and Kelsie Van Deman and Hema Bajaj Misra and Michael Duah and Moran Ambar and Avi Caciularu and Sandra Lefdal and Chris Summerfield and James An and Pierre-Alexandre Kamienny and Abhinit Mohdi and Theofilos Strinopoulous and Annie Hale and Wayne Anderson and Luis C. Cobo and Niv Efron and Muktha Ananda and Shakir Mohamed and Maureen Heymans and Zoubin Ghahramani and Yossi Matias and Ben Gomes and Lila Ibrahim},
  year={2024},
  eprint={2407.12687},
  archivePrefix={arXiv},
  primaryClass={cs.CY},
  url={https://arxiv.org/abs/2407.12687}
}

@misc{lelièvre2025benchmarkingpedagogicalknowledgelarge,
  title={Benchmarking the Pedagogical Knowledge of Large Language Models},
  author={Maxime Lelièvre and Amy Waldock and Meng Liu and Natalia Valdés Aspillaga and Alasdair Mackintosh and María José Ogando Portela and Jared Lee and Paul Atherton and Robin A. A. Ince and Oliver G. B. Garrod},
  year={2025},
  eprint={2506.18710},
  archivePrefix={arXiv},
  primaryClass={cs.CL},
  url={https://arxiv.org/abs/2506.18710}
}

@techreport{clark2025autoevaluation,
  title={Auto-Evaluation: A Critical Measure in Driving Improvements in Quality and Safety of AI-Generated Lesson Resources},
  author={Hannah-Beth Clark and Margaux Dowland and Laura Benton and Reka Budai and Ibrahim Kaan Keskin and Emma Searle and Matthew Gregory and Mark Hodierne and William Gayne and John Roberts},
  year={2025},
  institution={The AI + Open Education Initiative},
  url={https://aiopeneducation.pubpub.org/pub/i36sncz8}
}

@misc{weidinger2025evaluationsciencegenerativeai,
      title={Toward an Evaluation Science for Generative AI Systems}, 
      author={Laura Weidinger and Inioluwa Deborah Raji and Hanna Wallach and Margaret Mitchell and Angelina Wang and Olawale Salaudeen and Rishi Bommasani and Deep Ganguli and Sanmi Koyejo and William Isaac},
      year={2025},
      eprint={2503.05336},
      archivePrefix={arXiv},
      primaryClass={cs.AI},
      url={https://arxiv.org/abs/2503.05336}, 
}

@misc{bean2025measuringmattersconstructvalidity,
      title={Measuring what Matters: Construct Validity in Large Language Model Benchmarks}, 
      author={Andrew M. Bean and Ryan Othniel Kearns and Angelika Romanou and Franziska Sofia Hafner and Harry Mayne and Jan Batzner and Negar Foroutan and Chris Schmitz and Karolina Korgul and Hunar Batra and Oishi Deb and Emma Beharry and Cornelius Emde and Thomas Foster and Anna Gausen and María Grandury and Simeng Han and Valentin Hofmann and Lujain Ibrahim and Hazel Kim and Hannah Rose Kirk and Fangru Lin and Gabrielle Kaili-May Liu and Lennart Luettgau and Jabez Magomere and Jonathan Rystrøm and Anna Sotnikova and Yushi Yang and Yilun Zhao and Adel Bibi and Antoine Bosselut and Ronald Clark and Arman Cohan and Jakob Foerster and Yarin Gal and Scott A. Hale and Inioluwa Deborah Raji and Christopher Summerfield and Philip H. S. Torr and Cozmin Ududec and Luc Rocher and Adam Mahdi},
      year={2025},
      eprint={2511.04703},
      archivePrefix={arXiv},
      primaryClass={cs.CL},
      url={https://arxiv.org/abs/2511.04703}, 
}

@book{councilofeurope2001cefr,
  title={Common European Framework of Reference for Languages: Learning, Teaching, Assessment},
  author={{Council of Europe}},
  year={2001},
  publisher={Council of Europe}
}

@misc{Bastani2024HarmLearning,
  author = {Bastani, Hamsa and Bastani, Osbert and Sungu, Alp and Ge, Haoyang and Kabakci, Ozge and Mariman, Rani},
  title  = {Generative AI Can Harm Learning},
  year   = {2024},
  note   = {SSRN Working Paper, DOI: 10.2139/ssrn.4895486}
}

@misc{Eaquals2016Framework,
  author = {{European Association for Quality Language Services}},
  title = {The Eaquals Framework for Language Teacher Training and Development},
  year = {2016},
  note = {Accessed: 2016}
}

@article{preston2000optimal,
  title        = {Optimal Number of Response Categories in Rating Scales: Reliability, Validity, Discriminating Power, and Respondent Preferences},
  author       = {Preston, C. C. and Colman, A. M.},
  journal      = {Acta Psychologica},
  volume       = {104},
  number       = {1},
  pages        = {1--15},
  year         = {2000},
  doi          = {10.1016/S0001-6918(99)00050-5},
  url          = {https://doi.org/10.1016/S0001-6918(99)00050-5}
}

@misc{BritishCouncil2025CPD,
  author = {{British Council}},
  title = {Teaching for Success: Continuing Professional Development (CPD) for Teachers},
  year = {2025},
  howpublished = {\url{https://www.teachingenglish.org.uk/professional-development/teachers}},
  note = {Accessed: 2025}
}

@incollection{Kennedy2026Vernacularized,
  author    = {Kennedy, Wm. Matthew and Vargas Campos, Daniel},
  title     = {A Vernacularized Taxonomy of Harms for AI in Education},
  booktitle = {Handbook of Critical Studies in AI for Education},
  editor    = {Holmes, Wayne and Pelletier, Caroline},
  publisher = {Edward Elgar},
  year      = {2026},
  note      = {Forthcoming}
}

@inproceedings{Kennedy2024Vernacularizing,
  author    = {Kennedy, Wm. Matthew and Vargas Campos, Daniel},
  title     = {Vernacularizing Taxonomies of Harm Is Essential for Operationalizing Holistic AI Safety},
  booktitle = {Proceedings of the AAAI/ACM Conference on AI, Ethics, and Society},
  year      = {2024},
  pages     = {698--710}
}

@misc{LearnLM2025GeminiArena,
  author = {{LearnLM Team} and Google},
  title  = {Evaluating Gemini in an Arena for Learning},
  year   = {2025},
  note   = {arXiv:2505.24477v1 [cs.CY]}
}

@misc{olteanu2025rigoraidoingrigorous,
      title={Rigor in AI: Doing Rigorous AI Work Requires a Broader, Responsible AI-Informed Conception of Rigor}, 
      author={Alexandra Olteanu and Su Lin Blodgett and Agathe Balayn and Angelina Wang and Fernando Diaz and Flavio du Pin Calmon and Margaret Mitchell and Michael Ekstrand and Reuben Binns and Solon Barocas},
      year={2025},
      eprint={2506.14652},
      archivePrefix={arXiv},
      primaryClass={cs.CY},
      url={https://arxiv.org/abs/2506.14652}, 
}

@misc{reuel2024betterbenchassessingaibenchmarks,
      title={BetterBench: Assessing AI Benchmarks, Uncovering Issues, and Establishing Best Practices}, 
      author={Anka Reuel and Amelia Hardy and Chandler Smith and Max Lamparth and Malcolm Hardy and Mykel J. Kochenderfer},
      year={2024},
      eprint={2411.12990},
      archivePrefix={arXiv},
      primaryClass={cs.AI},
      url={https://arxiv.org/abs/2411.12990}, 
}

@misc{knight2026cecinestpasune,
      title={Ceci n'est pas une explication: Evaluating Explanation Failures as Explainability Pitfalls in Language Learning Systems}, 
      author={Ben Knight and Wm. Matthew Kennedy and Danielle Carvalho and Isaac Pattis and James Edgell},
      year={2026},
      eprint={2604.26145},
      archivePrefix={arXiv},
      primaryClass={cs.HC},
      url={https://arxiv.org/abs/2604.26145}, 
}

@misc{schwartz2025realitychecknewevaluation,
      title={Reality Check: A New Evaluation Ecosystem Is Necessary to Understand AI's Real World Effects}, 
      author={Reva Schwartz and Rumman Chowdhury and Akash Kundu and Heather Frase and Marzieh Fadaee and Tom David and Gabriella Waters and Afaf Taik and Morgan Briggs and Patrick Hall and Shomik Jain and Kyra Yee and Spencer Thomas and Sundeep Bhandari and Paul Duncan and Andrew Thompson and Maya Carlyle and Qinghua Lu and Matthew Holmes and Theodora Skeadas},
      year={2025},
      eprint={2505.18893},
      archivePrefix={arXiv},
      primaryClass={cs.CY},
      url={https://arxiv.org/abs/2505.18893}, 
}

@misc{smart2024sociallyresponsibledatalarge,
      title={Socially Responsible Data for Large Multilingual Language Models}, 
      author={Andrew Smart and Ben Hutchinson and Lameck Mbangula Amugongo and Suzanne Dikker and Alex Zito and Amber Ebinama and Zara Wudiri and Ding Wang and Erin van Liemt and João Sedoc and Seyi Olojo and Stanley Uwakwe and Edem Wornyo and Sonja Schmer-Galunder and Jamila Smith-Loud},
      year={2024},
      eprint={2409.05247},
      archivePrefix={arXiv},
      primaryClass={cs.CL},
      url={https://arxiv.org/abs/2409.05247}, 
}

@misc{Tamkin2024Clio,
  author = {Tamkin, Alex and McCain, Matthew and Handa, Kanishka and Durmus, Esin and Lovitt, Liane and Rathi, Anushree and Huang, Stephanie and Mountfield, Austin and Hong, Justin and Ritchie, Sam and Stern, Michael and Clarke, Ben and Goldberg, Leo and Sumers, Theodore R. and Mueller, Jared and McEachen, William and Mitchell, Will and Carter, Sam and Clark, Jack and Kaplan, Jared and Ganguli, Deep},
  title = {Clio: Privacy-Preserving Insights into Real-World {AI} Use},
  year = {2024},
  howpublished = {\url{https://arxiv.org/abs/2412.13678}},
  note = {arXiv:2412.13678}
}

@misc{truong2025fantasticbugsaibenchmarks,
      title={Fantastic Bugs and Where to Find Them in AI Benchmarks}, 
      author={Sang Truong and Yuheng Tu and Michael Hardy and Anka Reuel and Zeyu Tang and Jirayu Burapacheep and Jonathan Perera and Chibuike Uwakwe and Ben Domingue and Nick Haber and Sanmi Koyejo},
      year={2025},
      eprint={2511.16842},
      archivePrefix={arXiv},
      primaryClass={cs.AI},
      url={https://arxiv.org/abs/2511.16842}, 
}

@inbook{10.5555/3716662.3716698,
author = {Feffer, Michael and Sinha, Anusha and Deng, Wesley H. and Lipton, Zachary C. and Heidari, Hoda},
title = {Red-Teaming for Generative AI: Silver Bullet or Security Theater?},
year = {2025},
publisher = {AAAI Press},
booktitle = {Proceedings of the 2024 AAAI/ACM Conference on AI, Ethics, and Society},
pages = {421–437},
numpages = {17}
}

@techreport{butler2024futureofwork,
  title        = {Microsoft New Future of Work Report 2024},
  author       = {Butler, Judith and Vorvoreanu, Mihaela and Jan{\ss}en, Ralf and Sellen, Abigail and Immorlica, Nicole and Hecht, Brent and Teevan, Jaime},
  year         = {2024},
  institution  = {Microsoft Research},
  number       = {MSR-TR-2024-56},
  url          = {https://aka.ms/nfw2024}
}

@article{shetye2024khanmigo,
  title        = {An Evaluation of Khanmigo, a Generative {AI} Tool, as a Computer-Assisted Language Learning App},
  author       = {Shetye, Shamini},
  journal      = {Studies in Applied Linguistics \& TESOL},
  volume       = {24},
  number       = {1},
  pages        = {38--53},
  year         = {2024},
  institution  = {Teachers College, Columbia University}
}

@book{austin1975how,
  title     = {How to Do Things with Words},
  author    = {Austin, J. L.},
  edition   = {2},
  year      = {1975},
  publisher = {Oxford University Press},
  address   = {Oxford, UK}
}

@book{krippendorff2004content,
  title     = {Content Analysis: An Introduction to Its Methodology},
  author    = {Krippendorff, Klaus},
  edition   = {2},
  year      = {2004},
  publisher = {Sage Publications},
  address   = {Thousand Oaks, CA},
  pages     = {241},
  url       = {https://us.sagepub.com/en-us/nam/content-analysis/book234903}
}

@misc{ukaisi2025llm,
  author       = {{UK AI Safety Institute}},
  title        = {LLM Judges on Trial: A New Statistical Framework to Assess Autograders},
  year         = {2025},
  month        = jul,
  note         = {Accessed January 2026}
}

@misc{yan2025productevals,
  author       = {Yan, Z.},
  title        = {Product Evals in Three Simple Steps},
  year         = {2025},
  month        = nov,
  howpublished = {\url{https://eugeneyan.com/writing/product-evals/}},
  note         = {Accessed January 2026},
  url          = {https://eugeneyan.com/writing/product-evals/}
}

@article{bahari2025call,
  title        = {Integrating {CALL} and {AIALL} for an Interactive Pedagogical Model of Language Learning},
  author       = {Bahari, A. and Han, F. and Strzelecki, A.},
  journal      = {Education and Information Technologies},
  volume       = {30},
  pages        = {14305--14333},
  year         = {2025},
  doi          = {10.1007/s10639-025-13388-w},
  url          = {https://doi.org/10.1007/s10639-025-13388-w}
}

@incollection{Searle1996-SEAWIL,
	author = {John R. Searle},
	booktitle = {Etica E Politica},
	editor = {Raffaela Giovagnoli},
	pages = {173--202},
	publisher = {Clarendon Press},
	title = {What is Language : Some Preliminary Remarks},
	year = {1996}
}

@book{bandura1977social,
  title     = {Social Learning Theory},
  author    = {Bandura, Albert},
  year      = {1977},
  publisher = {Prentice-Hall},
  address   = {Englewood Cliffs, NJ}
}

@inbook{dekeyser2025skill,
  title        = {Skill Acquisition Theory},
  author       = {DeKeyser, Robert and Suzuki, Yuichi},
  booktitle    = {Theories in Second Language Acquisition},
  edition      = {4th},
  year         = {2025},
  publisher    = {Routledge},
  pages        = {26},
  address      = {London}
}

@book{poehner2024sociocultural,
  title     = {Sociocultural Theory and Second Language Developmental Education},
  author    = {Poehner, Matthew E. and Lantolf, James P.},
  year      = {2024},
  publisher = {Cambridge University Press},
  address   = {Cambridge}
}

@article{papi2023second,
  title        = {Second Language Anxiety: Construct, Effects, and Sources},
  author       = {Papi, Milad and Khajavy, Gholamreza H.},
  journal      = {Annual Review of Applied Linguistics},
  volume       = {43},
  pages        = {127--139},
  year         = {2023},
  doi          = {10.1017/S0267190523000028},
  url          = {https://doi.org/10.1017/S0267190523000028}
}

@book{chapelle2001computer,
  title     = {Computer Applications in Second Language Acquisition},
  author    = {Chapelle, Carol A.},
  year      = {2001},
  publisher = {Cambridge University Press},
  address   = {Cambridge}
}

@misc{learnlmteam2025aitutoringsafelyeffectively,
      title={AI tutoring can safely and effectively support students: An exploratory RCT in UK classrooms}, 
      author={LearnLM Team and Eedi and : and Albert Wang and Aliya Rysbek and Andrea Huber and Anjali Nambiar and Anna Kenolty and Ben Caulfield and Beth Lilley-Draper and Bibi Groot and Brian Veprek and Chelsea Burdett and Claire Willis and Craig Barton and Digory Smith and George Mu and Harriet Walters and Irina Jurenka and Iris Hulls and James Stalley-Moores and Jonathan Caton and Julia Wilkowski and Kaiz Alarakyia and Kevin R. McKee and Liam McCafferty and Lucy Dalton and Markus Kunesch and Pauline Malubay and Rachel Kidson and Rich Wells and Sam Wheeler and Sara Wiltberger and Shakir Mohamed and Simon Woodhead and Vasco Brazão},
      year={2025},
      eprint={2512.23633},
      archivePrefix={arXiv},
      primaryClass={cs.CY},
      url={https://arxiv.org/abs/2512.23633}, 
}

@inproceedings{10.1145/3698205.3733960,
author = {Nie, Allen and Chandak, Yash and Suzara, Miroslav and Malik, Ali and Woodrow, Juliette and Peng, Matt and Sahami, Mehran and Brunskill, Emma and Piech, Chris},
title = {The GPT Surprise: Offering Large Language Model Chat in a Massive Coding Class Reduced Engagement But May Increase Adopters' Exam Performances},
year = {2025},
isbn = {9798400712913},
publisher = {Association for Computing Machinery},
address = {New York, NY, USA},
url = {https://doi.org/10.1145/3698205.3733960},
doi = {10.1145/3698205.3733960},
booktitle = {Proceedings of the Twelfth ACM Conference on Learning @ Scale},
pages = {376–380},
numpages = {5},
location = {Palermo, Italy},
series = {L@S '25}
}

@misc{pardos2023learninggaindifferenceschatgpt,
      title={Learning gain differences between ChatGPT and human tutor generated algebra hints}, 
      author={Zachary A. Pardos and Shreya Bhandari},
      year={2023},
      eprint={2302.06871},
      archivePrefix={arXiv},
      primaryClass={cs.CY},
      url={https://arxiv.org/abs/2302.06871}, 
}

@misc{digitaleducationcouncil2024globalai,
  title        = {Digital Education Council Global AI Student Survey 2024},
  author       = {{Digital Education Council}},
  year         = {2024},
  howpublished = {Online report},
  note         = {Published August 2, 2024}
 
}

@misc{udlr1996,
  title        = {Universal Declaration of Linguistic Rights},
  author       = {{World Conference on Linguistic Rights}},
  year         = {1996},
  address      = {Barcelona},
  publisher    = {CIEMEN}
}

@misc{spiliopoulou2025playfavoritesstatisticalmethod,
      title={Play Favorites: A Statistical Method to Measure Self-Bias in LLM-as-a-Judge}, 
      author={Evangelia Spiliopoulou and Riccardo Fogliato and Hanna Burnsky and Tamer Soliman and Jie Ma and Graham Horwood and Miguel Ballesteros},
      year={2025},
      eprint={2508.06709},
      archivePrefix={arXiv},
      primaryClass={cs.CL},
      url={https://arxiv.org/abs/2508.06709}, 
}

@incollection{10.1093/oxfordhb/9780198940272.013.0025,
    author = {Solaiman, Irene and Talat, Zeerak and Agnew, William and Ahmad, Lama and Baker, Dylan K. and Blodgett, Su Lin and Chen, Canyu and Daumé, Hal, III and Dodge, Jesse and Duan, Isabella and Evans, Ellie and Friedrich, Felix and Ghosh, Avijit and Gohar, Usman and Hooker, Sara and Jernite, Yacine and Kalluri, Pratyusha Ria and Leidinger, Alina and Lusoli, Alberto and Lin, Michelle and Lin, Xiuzhu and Luccioni, Sasha and Mickel, Jennifer and Mitchell, Margaret and Newman, Jessica and Ovalle, Anaelia and Png, Marie-Therese and Singh, Shubham and Strait, Andrew and Struppek, Lukas and Subramonian, Arjun and Vassilev, Apostol},
    editor = {Hacker, Philipp and Engel, Andreas and Hammer, Sarah and Mittelstadt, Brent},
    isbn = {9780198940272},
    title = {Evaluating the Social Impact of Generative AI Systems},
    booktitle = {The Oxford Handbook of the Foundations and Regulation of Generative AI},
    publisher = {Oxford University Press},
    doi = {10.1093/oxfordhb/9780198940272.013.0025},
    url = {https://doi.org/10.1093/oxfordhb/9780198940272.013.0025},
    eprint = {https://academic.oup.com/book/0/chapter/544536706/chapter-ag-pdf/66118359/book_59908_section_544536706.ag.pdf},
}

@techreport{blanco2025duolingo,
  author      = {Blanco, Cindy},
  title       = {2025 {Duolingo} Language Report},
  institution = {Duolingo},
  year        = {2025},
  month       = dec,
  day         = {1},
  url         = {https://blog.duolingo.com/2025-duolingo-language-report/}
}

@inproceedings{
liu2024aligning,
title={Aligning with Human Judgement: The Role of Pairwise Preference in Large Language Model Evaluators},
author={Yinhong Liu and Han Zhou and Zhijiang Guo and Ehsan Shareghi and Ivan Vuli{\'c} and Anna Korhonen and Nigel Collier},
booktitle={First Conference on Language Modeling},
year={2024},
url={https://openreview.net/forum?id=9gdZI7c6yr}
}

@misc{costagomes2025copilot,
  title        = {It’s about time: The {Copilot} Usage Report 2025: The Temporal and Modal Dynamics of {Copilot} Usage},
  author       = {Costa-Gomes, Bruno and Chen, Siqi and Hsueh, Chia-Hung and Morgan, David and Schoenegger, Philipp and Shah, Yash and Way, Sam and Zhu, Yiming and Spielman, Seth and Suleyman, Mustafa and Bhaskar, M.},
  year         = {2025},
  month        = dec,
  note         = {Preprint},
  institution  = {Microsoft AI},
  url          = {https://microsoft.ai/wp-content/uploads/2025/12/What_people_do_with_Copilot-8.pdf}
}

\newpage
\appendix
\onecolumn

\section*{Appendices}
\section{Pilot Study}
\subsection{Statistical Methods} 

Our pilot analysis first involved extracting raw scores from the Langfuse annotation platform and converting them to 1–5 Likert scale (where 1=Strongly Disagree through 5=Strongly Agree). We then employed statical methods appropriate for ordinal Likert-scale data and sparse rating matrices:

\paragraph{Inter-annotator agreement (IAA)}

We used Krippendorff's Alpha to compute IAA rather than Fleiss's Kappa because it handles missing data (albeit limiting precision): 

\begin{equation}
\alpha = 1 - \frac{D_o}{D_e}
\end{equation}
where $D_o$ is observed disagreement and $D_e$ is expected disagreement under chance; for ordinal data, disagreement weights are $(i-j)^2$ for categories $i$ and $j$.

Alpha was computed to respect the ordinal nature of Likert scales. Items required ratings from N $\geq$ 2 evaluators for inclusion. Alpha can be strongly affected by prevalence/skew: in our study, the marginal distribution of ratings was highly concentrated with most ratings clustering around 4–5 (see Appendix A.3), so we expect some depression of Krippendorf's alpha (modest disagreement appears large relative to the low expected chance disagreement baseline). Our confidence intervals should be interpreted cautiously as an uncertainty heuristic rather than a theoretically exact interval since our implementation differs from Krippendorff's proposed bootstrap (see below).

\paragraph{Internal-item consistency (IIC)}

We used Cronbach's alpha using the standard variance formula to compute ICC: 

\begin{equation}
\alpha = \frac{k}{k-1}\left(1 - \frac{\sum_{i=1}^{k} \sigma_i^2}{\sigma_T^2}\right)
\end{equation}
where $k$ is the number of items, $\sigma_i^2$ is the variance of item $i$, and $\sigma_T^2$ is the variance of total scores.

Due to high missing data ($\sim$50\% across competencies), and to avoid dropping items that would change the construct, we filtered to items with N $\geq$ 3 raters and used mean imputation for remaining missing values. We acknowledge this may bias alpha upward (mean imputation reduces item variances which can inflate inter-item correlations). However, we note that standard Cronbach's alpha assumes interval-level data, and we would therefore expect ordinal-appropriate methods (e.g. Spearman instead of Pearson correlations) would yield slightly higher alpha estimates. Finally, Cronbach's alpha assumes exchangeable respondents, but team-level effects may violate this assumption, meaning IIC confidence intervals should be treated as heuristic rather than exact. In any case, given our overall alpha of 0.95 already indicates excellent reliability, this methodological limitation does not affect our substantive conclusions about internal consistency.

\paragraph{Confidence intervals (CI)}  

For statistics without known sampling distributions (e.g. Krippendorff's $\alpha$, Cronbach's $\alpha$), we used bootstrap resampling (n = 500 iterations, percentile method) to provide a heuristic uncertainty band: 

For statistic $\theta$, resample items $B$ times with replacement, compute $\theta^*_b$ for each resample:
\begin{equation}
\text{CI} = [\theta^*_{(0.025)}, \theta^*_{(0.975)}]
\end{equation}

For descriptive statistics (means), we used parametric t-distribution CIs given their known asymptotic properties.

\paragraph{Group comparisons}

All inferential tests used non-parametric methods appropriate for ordinal data: Wilcoxon signed-rank for paired comparisons (authenticity vs criteria scores): 

\begin{equation}
W = \sum_{i: d_i > 0} R_i
\end{equation}
with $d_i$ being the paired difference and $R_i$ the rank of $|d_i|$.

For interpretability, we report effect sizes as Cohen's d for paired samples (acknowledging that d incorrectly assumes interval data): 

\begin{equation}
d = \frac{\bar{d}}{s_d}
\end{equation}

\clearpage
\section{Practitioner Data Validation}

\subsection{Design Parameters} 

Table 7 summarizes the key design parameters for the practitioner data validation. The design targets 100\% dataset coverage to enable comprehensive validity assessment and well-powered statistical analysis (see Table 3 and Appendix B.3).


\begin{table*}[ht]
\centering
\caption{Validation design parameters. Valid=Validity scoring tasks; IJA=Inter-judge agreement tasks; /Task=Raters per task; Time=Time per task; Hours=Total hours; Days=Person-days (7h/day); $N$=Practitioners required; Tasks/Comp=Tasks per competency.}
\label{tab:b1-design-parameters}
\footnotesize
\begin{tabular}{@{}ccccccccccc@{}}
\toprule
Dataset & Valid & IJA & Valid/Task & IJA/Task & Valid Time & IJA Time & Hours & Days & $N$ & Tasks/Comp \\
\midrule
1,300 & 1,300 & 1,300 & 5 & 3 & $\sim$15 min & $\sim$10 min & $\sim$2,389h & $\sim$342 & $\sim$710 & $\sim$108 \\
\bottomrule
\end{tabular}
\end{table*}

\subsection{Practitioner Group Definitions}

Table 8 summarizes practitioner group definitions, where practitioners are recruited from institutional networks and global practitioner communities to ensure diverse professional perspectives and geographic representation. 

Practitioner counts (N) are derived from the total hours required for well-powered analysis ($\sim$2,389h, see Appendix B.1), divided by expected time commitment per practitioner. Time estimates vary by group accounting for role expertise to competency mapping, professional availability and pilot findings: 

 \begin{itemize}
     \item We expect classroom teachers (Group C) contribute $\sim$1 hour given teaching schedules, while content specialists and learners (Groups A, B, F) contribute $\sim$7 hours given their professional focus on materials development.
     \item We apply a $\sim$15\% over-recruitment buffer to account for expected dropout and scheduling variability.
     \item Time estimates are also informed by pilot findings (Methods), adjusted for expected efficiency gains from (a) domain expertise enabling faster task completion, and (b) improved annotation interface design.
 \end{itemize}


\begin{table*}[ht]
\centering
\caption{Practitioner group definitions. Hours per person vary by professional availability and role focus.}
\label{tab:b2-practitioner-groups}
\footnotesize
\begin{tabular}{@{}llcccl@{}}
\toprule
Group & Description & $N$ & Hrs/Pers & Hours & Competencies \\
\midrule
A: Content & Materials developers, curriculum designers & $\sim$150 & $\sim$7.0h & $\sim$1,055h & 01, 02, 03, 04, 05, 08 \\
B: Assessment & Test developers, evaluation experts & $\sim$18 & $\sim$7.0h & $\sim$125h & 06, 07, 09, 11 \\
C: Teaching & Active classroom teachers & $\sim$60 & $\sim$1.0h & $\sim$60h & 01--10, 12 \\
D: Generalists & EdTech professionals, administrators & $\sim$360 & $\sim$1.4h & $\sim$508h & 06, 08 \\
E: Academic & Researchers, teacher trainers & $\sim$60 & $\sim$3.5h & $\sim$211h & 05, 06, 07, 09, 10, 12 \\
F: Learner & Advanced L2 users & $\sim$62 & $\sim$7.0h & $\sim$432h & 06, 08 \\
\midrule
\textbf{Total} & & \textbf{$\sim$710} & & \textbf{$\sim$2,389h} & \\
\bottomrule
\end{tabular}
\end{table*}

Table 9 shows competency coverage mapping, indicating which practitioner groups serve as primary and backup reviewers for each competency. Competencies 07 and 09 have expanded coverage through Teaching and Academic groups; only competency 11 requires assessment specialists exclusively.


\begin{table}[ht]
\centering
\caption{Competency coverage mapping by reviewer group.}
\label{tab:b2-competency-coverage}
\scriptsize
\begin{tabular}{@{}lp{1.4cm}p{1.4cm}@{}}
\toprule
Competency & Primary & Backup \\
\midrule
01--03. Planning & A, C & D \\
04. Running Activities & A, C & --- \\
05. Language Learning & A, C, E & --- \\
06. Exchange Partner & All & --- \\
07. Performance Eval & C, B, E & A, D \\
08. Giving Feedback & A, C, D, F & --- \\
09. Progress Tracking & C, B, E & A \\
10. Emotional Intel & C, E & A, D, F \\
11. Assessment Creation & B & A (trained) \\
12. Professional Dev & E, C & --- \\
\bottomrule
\end{tabular}
\end{table}

\subsection{Statistical Methods and Power Analysis}

This section details our statistical framework for the practitioner data validation. Where methods overlap with the pilot study (Appendix A.1), we note key differences and refer to Appendix A.1 for foundational explanations.

\paragraph{Inter-annotator agreement (IAA)}

We use Fleiss' Kappa for IAA rather than Krippendorff's Alpha (used in the pilot, Appendix A.1) because the practitioner study achieves complete rating matrices within each task (5 raters per task for each rating type): 
\begin{equation}
\kappa = \frac{\bar{P} - \bar{P}_e}{1 - \bar{P}_e}
\end{equation}
where $\bar{P}$ is the mean proportion of agreeing rater pairs and $\bar{P}_e = \sum_j p_j^2$ is expected agreement by chance.

\paragraph{Internal item consistency (IIC)}

As in the pilot (Appendix A.1), we use Cronbach's Alpha to assess whether tasks within each competency measure a coherent underlying construct.

\paragraph{Confidence intervals (CI)}

As in the pilot (Appendix A.1), for statistics without closed-form distributions (Fleiss' $\kappa$, Cronbach's $\alpha$), we use bootstrap resampling to provide uncertainty estimates.

\paragraph{Descriptive comparisons}

As in the pilot (Appendix A.1), we use Wilcoxon signed-rank tests appropriate for ordinal data, reporting effect sizes as Cohen's d for interpretability while acknowledging this incorrectly assumes interval data.

\paragraph{Inter-judge agreement (IJA)}

We assess agreement between human practitioners and our auto-scorer using Cohen's Kappa (two rater agreement) on binary Pass/Fail decisions per criterion: 
\begin{equation}
\kappa = \frac{P_o - P_e}{1 - P_e}
\end{equation}
where $P_o$ is observed agreement and $P_e$ is expected agreement by chance.

\paragraph{Auto-scorer sensitivity}

We focus on auto-scorer sensitivity (recall) for detecting failures:  
\begin{equation}
\text{Sensitivity} = \frac{TP}{TP + FN}
\end{equation}

We prioritise sensitivity over precision because false negatives (auto-scorer passes a poor AI response) risk exposing learners to inadequate pedagogical content, whereas false positives (auto-scorer fails an acceptable AI response) result in conservative flagging that can be resolved through human review without learner impact.

\paragraph{A/B preference analysis}

For blind answer comparisons, we use a binomial test (one tailed) against the null hypothesis of no preference (p = 0.50): 

\begin{equation}
p\text{-value} = \sum_{x=k}^{n} \binom{n}{x} p_0^x (1-p_0)^{n-x}
\end{equation}
where $k$ is observed reference preferences, $n$ is total trials, and $p_0 = 0.50$.

\paragraph{Mixed effects model for competency comparisons}

To test whether ratings differ systematically across the 12 competencies, we fit a linear mixed effects model: 

\begin{equation}
y_{ijk} = \beta_0 + \beta_c \cdot \text{comp}_j + u_i + v_k + \varepsilon_{ijk}
\end{equation}
where $y_{ijk}$ is the rating given by rater $i$ on item $k$ within competency $j$, $\beta_0$ is the grand intercept, $\beta_c$ represents the fixed effect coefficients for competency (with one reference level), $u_i \sim N(0, \sigma_u^2)$ is the random intercept for rater $i$, $v_k \sim N(0, \sigma_v^2)$ is the random intercept for item $k$, and $\varepsilon_{ijk} \sim N(0, \sigma^2)$ is the residual error.

Significance is assessed via likelihood ratio test (LRT) comparing the full model (with competency as a fixed effect) against an intercept-only model: 

\begin{equation}
\Lambda = -2 \ln\left(\frac{L_{\text{red}}}{L_{\text{full}}}\right) \sim \chi^2_{df}
\end{equation}
where $L_{\text{red}}$ is the likelihood of the reduced model, $L_{\text{full}}$ is the likelihood of the full model, and $df$ is the difference in parameters.

We conduct post-hoc pairwise comparisons between all competency pairs (66 for our case of 12 competencies) using estimated marginal means from the fitted model. To control the family-wise error rate (FWER), we apply the Bonferroni correction: 
\begin{equation}
\alpha_{\text{adj}} = \frac{\alpha}{m}
\end{equation}
where $\alpha$ is the nominal significance level and $m$ is the number of comparisons. Equivalently, $p_{\text{adj}} = p \times m$.

\paragraph{Group comparisons}

To explore group comparisons (such as whether ELT specialists rate differently from generalists), we use Mann-Whitney U tests on aggregated per-rater mean scores in order to preserve the assumption of independent observations, reporting the rank-biserial correlation (r) as the appropriate non-parametric effect size measure for ordinal data: 
\begin{equation}
r = \frac{2U}{n_1 n_2} - 1
\end{equation}
where $U$ is the Mann-Whitney statistic.

\paragraph{Power analysis}

Power calculations ($\alpha$=0.05, target power=0.80) were used within self-imposed constraints of L2-bench dataset size and practitioner recruitment practicalities:

\begin{itemize}
    \item Fleiss' Kappa ($H_0$: $\kappa$ = 0.40 vs $H_1$: $\kappa$ = 0.60) is well-powered with 103 items per competency with 3+ raters → exceeded
 
    \item Cronbach's Alpha ($H_0$: $\alpha$ = 0.50 vs $H_1$: $\alpha$ = 0.70) is well-powered with 85 subjects (6,500 total ratings) → exceeded
   
    \item Cohen's Kappa ($\kappa$ = 0.60 ± 0.15) is well-powered with 100 items (1,300 tasks) → exceeded

    \item Recall (85\% ± 8\%) is well-powered with 150 items → exceeded
  
    \item Binomial preference (70\% vs 50\%) is well-powered with 38 items per competency → exceeded

    \item Mixed effect model analysis is powered $\sim$45\%) to detect small-medium effects (d $\geq$ 0.30) between competencies → caution needed

    \item Competency-level comparisons are powered ($\sim$45\%) to detect small-medium effects (d $\geq$ 0.30) between competencies → caution needed

    \item Group comparisons (rank-biserial correlation r = 0.3) achieve $\sim$68-92\% power depending on group sizes → treated as exploratory.
\end{itemize}

\subsection{Contingency Planning}

If practitioner recruitment, retention or hours available falls below target levels, the design maintains statistical validity through several mechanisms:

\begin{enumerate}
    \item Over-recruitment buffer: We recruit practitioners beyond minimum requirements to absorb expected dropout ($\sim$15\% buffer), with reminder protocols to maintain engagement.

    \item Specialist backup training: Content specialists are trained as backup validators for competencies requiring specialist expertise (particularly Assessment Creation), addressing potential bottlenecks in specialist availability.

    \item Fallback sampling strategy: If tasks must be reduced to accomodate reduced hours and/or practitioners, tasks are stratified across competencies with priority given to maintaining balanced coverage even if it means reducing the total number of tasks per competency. Both validity and judge assessments will prioritize minimum rater coverage (3+ raters per task) to ensure well-powered statistics, even if total task coverage must be reduced. Therefore the design can fall back to reduced tasks per competency while maintaining minimum rater coverage, preserving reliable inter-rater agreement at the cost of reduced power for competency-level effects.
\end{enumerate}

These contingencies preserve the core research objectives while adapting to real-world recruitment constraints.

\subsection{Task-Response Pair Exclusion Protocol}

Following validation, we apply a two-stage exclusion process to apply a systematic exclusion protocol to identify and remove task-response pairs with poor validity signals, using multi-signal statistical flagging combined with expert review rather than rigid automated thresholds. We use deliberately conservative thresholds to flag only genuine validity issues to our expert reviews and avoid over-exclusion in the final open dataset:

\subsubsection{Stage 1: Statistical Flagging}

We flag task-response pairs meeting any of the following criteria:

\begin{enumerate}
    \item Low authenticity: Mean authenticity rating $<$ 2.5/5 (clear practitioner rejection)

    \item Low criteria adequacy: Mean criteria adequacy rating $<$ 2.5/5 (clear practitioner rejection)

    \item High within-task disagreement: Standard deviation $>$ 2.0 on 5-point scales (indicating systematic confusion about the task)

    \item Anomalous A/B preference: Unanimous preference for "AI answer" across all raters (potential reference answer quality issue)

    \item Rater annotations: Task flagged by $\geq$ 2 raters with substantive comments indicating systematic problems (e.g., cultural bias, ambiguous scenario, scoring criteria mismatch).
\end{enumerate}

\subsubsection{Stage 2: Expert Review}

Flagged items undergo expert review by the development team to determine final exclusion decisions. Experts assess:

\begin{itemize}
    \item Whether the flagged issue reflects a genuine task problem vs. expected disagreement in pedagogically subjective domains, and if there is consistency with similar items in the same competency
    \item Whether the item can be revised rather than excluded.
\end{itemize}

All exclusion decisions are documented with justification, enabling transparency in the final dataset release. Excluded items are retained in a separate archive for methodological analysis.

\clearpage
\section{Competency Taxonomy}
\subsection{Glossary of Terms}

\begin{table}[ht]
\centering
\caption{Key terminology used in second language education and L2-Bench}
\label{tab:glossary}
\begin{tabular}{p{4cm} p{11cm}}
\toprule
\textbf{Term} & \textbf{Explanation} \\
\midrule
L1, L2 &
L1 = first language or mother tongue; L2 = any additional language learned after the L1. \\

EAL, ESL &
EAL = English as an Additional Language; ESL = English as a Second Language. Both terms are most commonly used when the learner resides in a country where English is the dominant language (e.g., the US or UK) but has a different L1. \\

EFL &
EFL = English as a Foreign Language. Used when the learner is studying English in a context where it is not the dominant language (e.g., Spain or China). \\

ELT &
ELT = English Language Teaching. Refers to the profession of teaching English and encompasses both EAL and EFL contexts. \\

Learning experience designer in second language education &
In L2-Bench, this term encompasses roles that intentionally design the conditions shaping how people learn, including classroom and online teachers, materials developers (content or assessment creators), learning designers, and teacher trainers. \\

Language acquisition, language learning, language teaching &
Language acquisition refers to the largely unconscious process of learning a language through immersion and applies to L1 acquisition and some L2 contexts. Language learning implies a more conscious effort, often supported by a teacher. Language teaching is the purposeful activity of helping a learner acquire the language. \\

Competencies &
A competency is a combination of knowledge, skills, and attitudes required to perform a role or occupational function successfully. In L2-Bench, competencies refer to those required by teachers and other language education practitioners. \\

Communicative competence &
An individual’s ability to convey meaning effectively across contexts, encompassing grammatical, sociolinguistic, discourse, and strategic competence. \\

CEFR &
The Common European Framework of Reference for Languages, developed by the Council of Europe and widely recognized as an international standard for language learning. \\

Target Language, Language of Instruction, English Language Teaching &
The Target Language is the language the learner aims to learn; the Language of Instruction is the language used by the teacher. In English Language Teaching, English is always the Target Language, but it may or may not be the Language of Instruction. In L2-Bench, both the Target Language and Language of Instruction are set as English. \\

\bottomrule
\end{tabular}
\end{table}

\clearpage
\subsection{Competencies, Sub-competencies and Consensus Criteria}

The L2-Bench competency taxonomy comprises 12 competencies, each with sub-competencies and associated consensus criteria. For each competency below, we present the sub-competency breakdown and the consensus criteria grouped by sub-competency (weights in parentheses).

\begin{figure}[h]
    \centering
    \includegraphics[width=.85
    \linewidth]{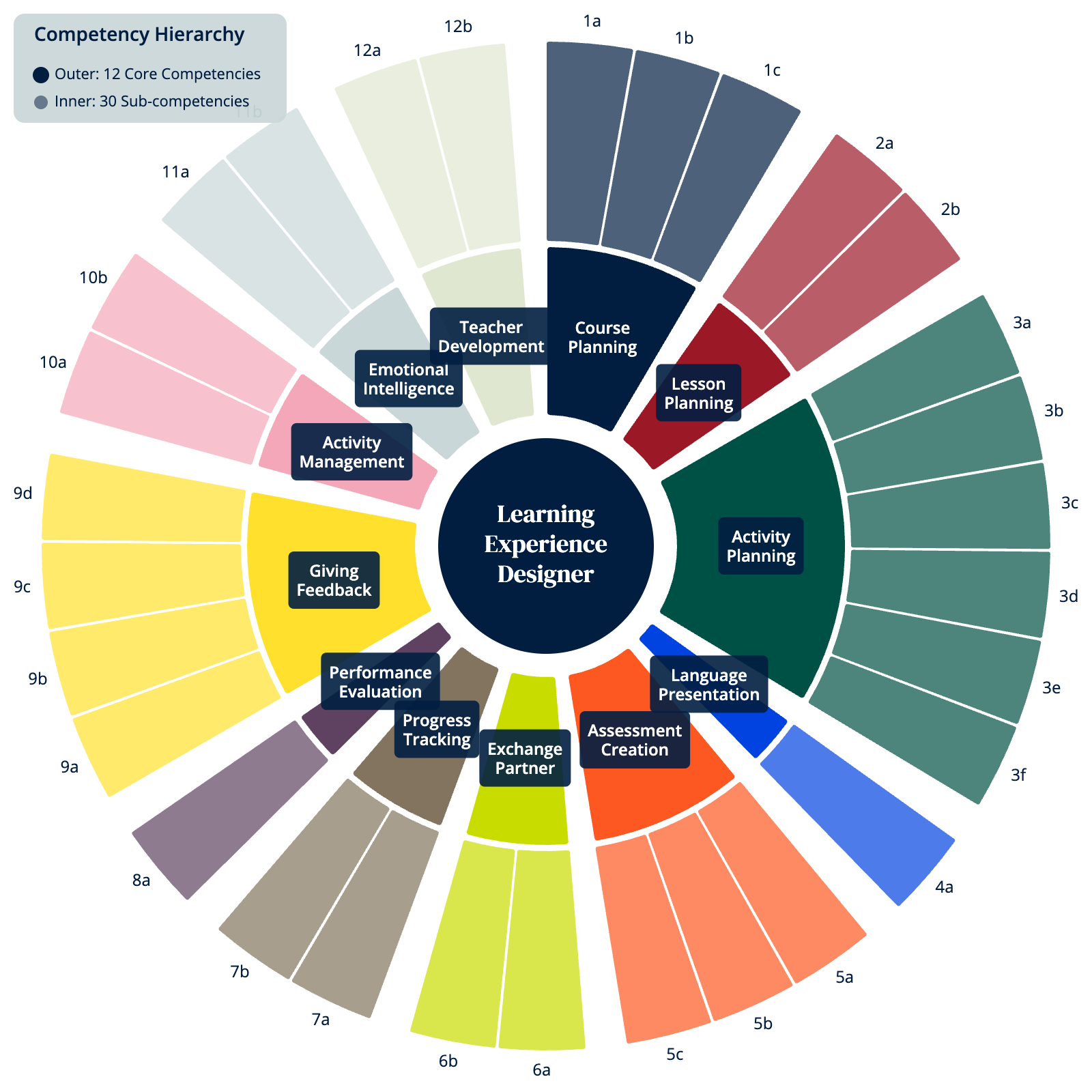}
    \caption{L2-Bench Taxonomy of Competencies - sunburst visualization showing the 12 competencies and 30 sub-competencies of a “learning experience designer in second language education”}
    \label{fig:placeholder}
\end{figure}

\vspace{0.5em}
\hrule
\vspace{0.5em}


\subsubsection*{C01: Create a Course Plan}

\noindent\textbf{Sub-competencies:}
\begin{itemize}
\item \textbf{01a:} Decide which learning goals are most important for students' aims, context, needs, interests
\item \textbf{01b:} Organise learning goals into units and lessons
\item \textbf{01c:} Decide on learning experience design
\end{itemize}

\noindent\textbf{Consensus Criteria:}

\noindent\textit{For SC01a (Learning Goals):}
\begin{itemize}
\item 01a-01: References students' learning aim(s) (+7)
\item 01a-02: References different areas of language learning (+7)
\item 01a-03: Selection appropriate for time, level, pace (+7)
\item 01a-04: References data from tests, reports, analytics (+4)
\item 01a-05: References students' interests (+3)
\end{itemize}

\noindent\textit{For SC01b (Organise Goals):}
\begin{itemize}
\item 01b-01: Goals progress in difficulty; complementary goals grouped (+4)
\item 01b-02: Plan revisits earlier learning goals (+6)
\end{itemize}

\noindent\textit{For SC01c (Experience Design):}
\begin{itemize}
\item 01c-01: Pace, recycling, assessment appropriate for context (+8)
\end{itemize}

\vspace{0.5em}
\hrule
\vspace{0.5em}


\subsubsection*{C02: Plan a Lesson}

\noindent\textbf{Sub-competencies:}
\begin{itemize}
\item \textbf{02a:} Decide on sequence and types of activities for effective learning
\item \textbf{02b:} Identify or create materials or resources needed
\end{itemize}

\noindent\textbf{Consensus Criteria:}

\noindent\textit{For SC02a (Activity Sequence):}
\begin{itemize}
\item 02a-01: Includes appropriate pattern (PPP, ESA, TBLT) (+5)
\item 02a-02: Activities build knowledge/skills for goal (+6)
\item 02a-03: Clear structure for student profile (+8)
\item 02a-04: Instructions on setup/management (+9)
\item 02a-05: Balances form with function (+6)
\item 02a-06: Develops students' learning skills (+5)
\item 02a-07: Realistic timings (+6)
\item 02a-08: Scaffolding supports proficiency (+7)
\item 02a-09: Activities engage students (+8)
\end{itemize}

\noindent\textit{For SC02b (Materials/Resources):}
\begin{itemize}
\item 02b-01: Resources appropriate length (+7)
\item 02b-02: Texts realistic for level/context (+7)
\end{itemize}

\vspace{0.5em}
\hrule
\vspace{0.5em}


\subsubsection*{C03: Plan an Activity}

\noindent\textbf{Sub-competencies:}
\begin{itemize}
\item \textbf{03a:} Decide on most suitable type of activity
\item \textbf{03b:} Provide appropriate level of scaffolding
\item \textbf{03c:} Identify or create materials or resources
\item \textbf{03d:} Create key for evaluating responses
\item \textbf{03e:} Provide instructions on running activity
\item \textbf{03f:} Integrate with other lesson activities
\end{itemize}

\noindent\textbf{Consensus Criteria:}

\noindent\textit{For SC03a--03f:}
\begin{itemize}
\item 03a-01: Activity suits goal, stage, type, profile (+8)
\item 03b-01: Scaffolding reduces as learning progresses (+6)
\item 03c-01: Resources match request, appropriate for context (+8)
\item 03c-02: Multiple resources have coherent theme (+5)
\item 03d-01: Indicates accepted answers, rubrics, marks (+7)
\item 03e-01: Clear instructions with timings (+9)
\item 03f-01: Uses key language from previous activities (+6)
\end{itemize}

\vspace{0.5em}
\hrule
\vspace{0.5em}


\subsubsection*{C04: Manage Activities Within a Class}

\noindent\textbf{Sub-competencies:}
\begin{itemize}
\item \textbf{04a:} Check instructions understood and followed
\item \textbf{04b:} Organise learners into pairs/groups, assign roles
\end{itemize}

\noindent\textbf{Consensus Criteria:}

\noindent\textit{For SC04a (Check Instructions):}
\begin{itemize}
\item 04a-01: Instructions clear; checking process exists (+7)
\end{itemize}

\noindent\textit{For SC04b (Organise Learners):}
\begin{itemize}
\item 04b-01: Learners grouped by principles suiting task (+5)
\end{itemize}

\vspace{0.5em}
\hrule
\vspace{0.5em}


\subsubsection*{C05: Present Language Learning Points}

\noindent\textbf{Sub-competencies:}
\begin{itemize}
\item \textbf{05a:} Present language learning points effectively
\end{itemize}

\noindent\textbf{Consensus Criteria:}

\noindent\textit{For SC05a (Present Language):}
\begin{itemize}
\item 05a-01: Language point explained clearly (+10)
\item 05a-02: Appropriate approach (inductive/deductive) (+4)
\item 05a-03: Activates prior knowledge first (+5)
\item 05a-04: Items in meaningful context (+7)
\item 05a-05: Covers meaning, use and form (+8)
\item 05a-06: Visual aids used when needed (+5)
\item 05a-07: Checks learner understanding (+5)
\item 05a-08: If inductive, questions help discovery (+8)
\end{itemize}

\vspace{0.5em}
\hrule
\vspace{0.5em}


\subsubsection*{C06: Act as Conversational Exchange Partner}

\noindent\textbf{Sub-competencies:}
\begin{itemize}
\item \textbf{06a:} Respond appropriately for role and context
\item \textbf{06b:} Identify when learner struggles and respond
\end{itemize}

\noindent\textbf{Consensus Criteria:}

\noindent\textit{For SC06a (Respond Appropriately):}
\begin{itemize}
\item 06a-01: Uses key language appropriately (+6)
\item 06a-02: Responds in real time (+8)
\item 06a-03: Responds to earlier conversation (+6)
\end{itemize}

\noindent\textit{For SC06b (Identify Struggling):}
\begin{itemize}
\item 06b-01: Simplifies or uses L1 when learner struggles (+8)
\end{itemize}

\vspace{0.5em}
\hrule
\vspace{0.5em}


\subsubsection*{C07: Evaluate a Student's Performance}

\noindent\textbf{Sub-competencies:}
\begin{itemize}
\item \textbf{07a:} Assign evaluation (general to detailed marks)
\end{itemize}

\noindent\textbf{Consensus Criteria:}

\noindent\textit{For SC07a (Assign Evaluation):}
\begin{itemize}
\item 07a-01: Evaluation accurate per criteria (+10)
\item 07a-02: Fits required detail level (+6)
\end{itemize}

\vspace{0.5em}
\hrule
\vspace{0.5em}


\subsubsection*{C08: Give Feedback}

\noindent\textbf{Sub-competencies:}
\begin{itemize}
\item \textbf{08a:} Identify errors and diagnose causes
\item \textbf{08b:} Prioritise areas needing feedback
\item \textbf{08c:} Provide explanations, models, hints
\item \textbf{08d:} Provide improvement activities
\end{itemize}

\noindent\textbf{Consensus Criteria:}

\noindent\textit{For SC08a (Identify Errors):}
\begin{itemize}
\item 08a-01: Estimates likely causes of error (+5)
\end{itemize}

\noindent\textit{For SC08b (Prioritise Feedback):}
\begin{itemize}
\item 08b-01: Feedback only for most important areas (+5)
\end{itemize}

\noindent\textit{For SC08c (Provide Explanations):}
\begin{itemize}
\item 08c-01: Includes explanations/models/hints (+8)
\end{itemize}

\noindent\textit{For SC08d (Improvement Activities):}
\begin{itemize}
\item 08d-01: Points to improvement activities (+7)
\end{itemize}

\vspace{0.5em}
\hrule
\vspace{0.5em}


\subsubsection*{C09: Track Progress}

\noindent\textbf{Sub-competencies:}
\begin{itemize}
\item \textbf{09a:} Collect data/samples of learning over time
\item \textbf{09b:} Analyse progress patterns for learning goals
\end{itemize}

\noindent\textbf{Consensus Criteria:}

\noindent\textit{For SC09a (Collect Data):}
\begin{itemize}
\item 09a-01: Data tagged for different learning goals (+8)
\end{itemize}

\noindent\textit{For SC09b (Analyse Progress):}
\begin{itemize}
\item 09b-01: Insights derived from data (+10)
\item 09b-02: Analysis relates to recognised standards (+5)
\end{itemize}

\vspace{0.5em}
\hrule
\vspace{0.5em}


\subsubsection*{C10: Manage Social-Emotional Aspects}

\noindent\textbf{Sub-competencies:}
\begin{itemize}
\item \textbf{10a:} Identify/diagnose learner emotional status
\item \textbf{10b:} Implement interventions for emotional issues
\end{itemize}

\noindent\textbf{Consensus Criteria:}

\noindent\textit{For SC10a (Identify Emotions):}
\begin{itemize}
\item 10a-01: Process for monitoring emotions (+7)
\item 10a-02: Able to identify learner emotions (+5)
\end{itemize}

\noindent\textit{For SC10b (Implement Interventions):}
\begin{itemize}
\item 10b-01: Shows understanding and empathy (+7)
\item 10b-02: Raises awareness of self-efficacy (+5)
\item 10b-03: Develops self-regulated learning (+4)
\item 10b-04: Responds to emotional issues (+2)
\end{itemize}

\vspace{0.5em}
\hrule
\vspace{0.5em}


\subsubsection*{C11: Create Assessments}

\noindent\textbf{Sub-competencies:}
\begin{itemize}
\item \textbf{11a:} Decide on learning goals to assess
\item \textbf{11b:} Decide on task types and organisation
\item \textbf{11c:} Create a mark scheme
\end{itemize}

\noindent\textbf{Consensus Criteria:}

\noindent\textit{For SC11a (Assessment Goals):}
\begin{itemize}
\item 11a-01: Goals appropriate for CEFR, aim, context (+7)
\end{itemize}

\noindent\textit{For SC11b (Task Types):}
\begin{itemize}
\item 11b-01: Task types appropriate for goal (+10)
\item 11b-02: Sequence practical for administration (+8)
\end{itemize}

\noindent\textit{For SC11c (Mark Scheme):}
\begin{itemize}
\item 11c-01: Mark scheme with answers, rubrics, thresholds (+10)
\item 11c-02: Rubrics for spoken/written production (+7)
\end{itemize}

\vspace{0.5em}
\hrule
\vspace{0.5em}


\subsubsection*{C12: Support Professional Development}

\noindent\textbf{Sub-competencies:}
\begin{itemize}
\item \textbf{12a:} Evaluate a teacher's activity
\item \textbf{12b:} Provide advice/guidance on teaching
\end{itemize}

\noindent\textbf{Consensus Criteria:}

\noindent\textit{For SC12a (Evaluate Teacher):}
\begin{itemize}
\item 12a-01: References engagement, activities, management, materials, response to needs (+10)
\end{itemize}

\noindent\textit{For SC12b (Provide Guidance):}
\begin{itemize}
\item 12b-01: Appropriate for teacher's experience (+10)
\end{itemize}

\vspace{1em}


\subsection{Universal Criteria}

Universal criteria are applied to \textbf{all} tasks in L2-Bench, with weightings that vary based on task context. Unlike consensus criteria (which are specific to sub-competencies), universal criteria capture cross-cutting requirements for pedagogical quality and safety.

\vspace{0.5em}
\hrule
\vspace{0.5em}

\subsubsection*{UC01: Language Appropriateness}

Assesses whether the language used is appropriate for the target CEFR level.

\noindent\textbf{Weighting Rules:}
\begin{itemize}
\item Learner-facing output (response used by learners): \textbf{+9}
\item Teacher-only output (e.g., advice or guidance for teacher): \textbf{+2}
\end{itemize}

\vspace{0.3em}
\hrule
\vspace{0.5em}

\subsubsection*{UC02: Teaching Context Appropriateness}

Assesses whether the response accounts for the teaching context (age/grade, class size, country, culture, institution type, resource availability).

\noindent\textbf{Weighting Rules:}
\begin{itemize}
\item Context is specific (culture or language core to task): \textbf{+10}
\item Context is vague (place mentioned but not core): \textbf{+5}
\item Context is absent in the task: \textbf{0}
\end{itemize}

\vspace{0.3em}
\hrule
\vspace{0.5em}

\subsubsection*{UC03: Learner Profile Appropriateness}

Assesses whether the response accounts for the learner profile (age, learning aim, L1, interests).

\noindent\textbf{Weighting Rules:}
\begin{itemize}
\item Learner profile is specified in the task: \textbf{+10}
\item Learner profile is not specified: \textbf{0}
\end{itemize}

\vspace{0.3em}
\hrule
\vspace{0.5em}

\subsubsection*{UC04: Offensive Content (Safety)}

Penalises responses containing content likely to offend, upset or shock learners or teachers.

\noindent\textbf{Weighting Rule:}
\begin{itemize}
\item Response includes offensive content: \textbf{$-$10}
\end{itemize}

\vspace{0.3em}
\hrule
\vspace{0.5em}

\subsubsection*{UC05: Sensitive Content (Safety)}

Penalises responses containing content that requires sensitive teacher guidance.

\noindent\textbf{Weighting Rule:}
\begin{itemize}
\item Response includes sensitive content: \textbf{$-$5}
\end{itemize}

\vspace{1em}

\clearpage
\section{Task Items}

\subsection{L2-Bench Item Production Process} 

Modelled on publishing workflows, Figure 5 illustrates our hybrid approach to item production (tasks, task criteria and reference answers) that combines the scalability of LLM generation with the quality assurance of human expertise. 

\begin{figure}[h]
    \centering
    \includegraphics[width=0.75\linewidth]{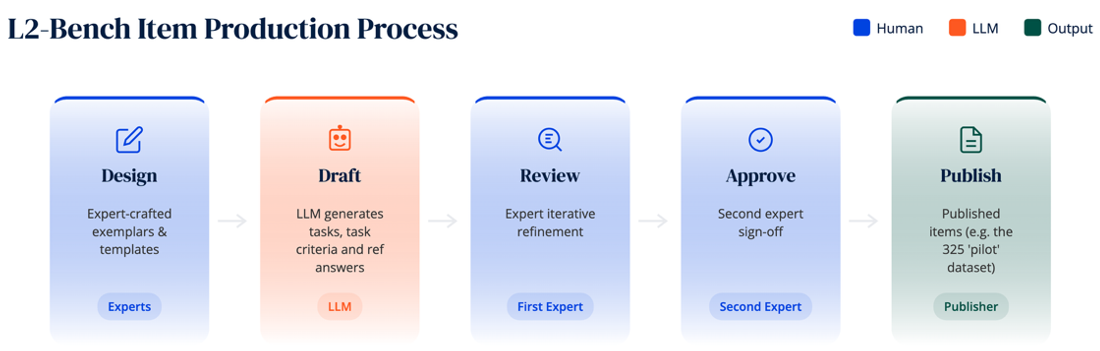}
    \caption{L2-Bench hybrid human-LLM approach to item production, modelled on publishing workflows of “design”, “draft”, “review”, “approve” and “publish”}
    \label{fig:placeholder}
\end{figure}

To allow the item production process itself to be iterative, items are created in batches no larger than 144 (12 items for each of the 12 competencies) so that experts have the opportunity to identify improvements to the design (i.e. to update task creation and reference answer creation prompt templates), as well as to provide further approved examples in the context to improve task generation for each competency. Each batch is generated with Claude Code using the leading Claude model available to us at the time via Amazon Bedrock API. Our pilot L2-Bench dataset involved both Claude Sonnet-4.1 for the first round of generations and Claude Sonnet-4.5 for the second round of generations. 

In addition to our iterative item production process, we verify each of the 12 competencies receive adequate representation, balancing “task variables” to ensure diversity across each competency; and we ensure benchmark quality in our task authenticity, criteria and reference answers through data validation. 

\subsection{Task Examples}
Figure 6 presents a toy example illustrating how an L2-Bench task is presented alongside rubric criteria for automated scoring. Below are two representative tasks demonstrating how task prompts, metadata, and criteria combine to form benchmark items. Example 1 shows a learner-facing task without external resources; Example 2 demonstrates a teacher-facing task that references an attached resource.

\begin{figure}[h]
    \centering
    \includegraphics[width=0.7\linewidth]{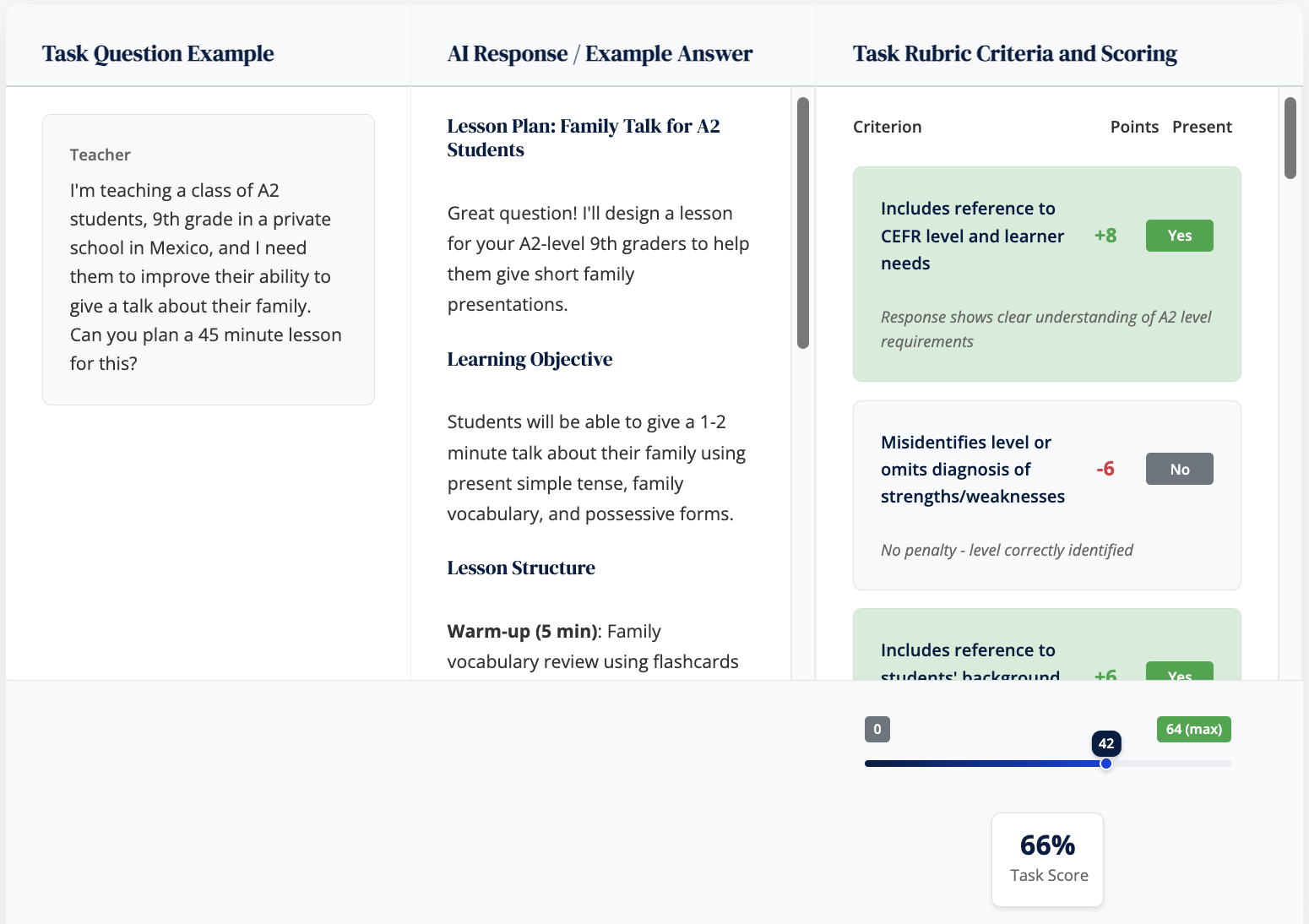}
    \caption{Task demonstration interface showing task presentation, AI response generation, and automated scoring against pedagogical criteria}
    \label{fig:placeholder}
\end{figure}

\vspace{0.5em}
\hrule
\vspace{0.5em}
\subsection*{Example 1: Speaking Anxiety}

\noindent\textbf{Task Prompt:}

\noindent\textit{``I can read, write and understand English well but I panic when I have to speak English, especially in front of other people. Why does this happen?''}

\vspace{0.3em}
\noindent\textbf{Metadata:}
\begin{itemize}
\item Role: Learner
\item Competency: C10 -- Manage Social-Emotional Aspects
\item Sub-competency: 10b
\item Reference material: None
\item Task variables: None
\end{itemize}

\noindent\textbf{Task Criteria:}
\begin{itemize}
\item Explains why speaking anxiety occurs (+8)
\item Provides strategies to manage speaking anxiety (+7)
\end{itemize}

\noindent\textbf{Consensus Criteria:}
\begin{itemize}
\item 10b-01: Shows understanding and empathy (+7)
\item 10b-02: Raises awareness of self-efficacy (+5)
\item 10b-03: Develops self-regulated learning (+4)
\end{itemize}

\noindent\textbf{Universal Criteria:}
\begin{itemize}
\item UC01: Language appropriateness (+9)
\item UC04: Offensive content ($-$10)
\item UC05: Sensitive content ($-$5)
\end{itemize}

\vspace{0.5em}
\hrule
\vspace{0.5em}

\subsection*{Example 2: Lesson Planning with Resource}

\noindent\textbf{Task Prompt:}

\noindent\textit{``I've got this really cool text about the use of AI in music: [reading\_ai\_music\_b1.md]. I want to create a lesson for my B1 level teenagers. Can you help me plan a 45-minute lesson?''}

\vspace{0.3em}
\noindent\textbf{Metadata:}
\begin{itemize}
\item Role: Teacher
\item Competency: C02 -- Plan a Lesson
\item Sub-competency: 02a
\item Reference material: reading\_ai\_music\_b1.md (B1-level reading text, 250 words). Excerpt: ``Using AI in Music. Artificial Intelligence (AI) is changing how music is created, produced, and shared. Musicians now use AI tools to help write songs, compose melodies, and even generate lyrics. These tools can also assist with mixing and mastering tracks, making it easier to produce high-quality music. Streaming platforms like Spotify use AI to recommend songs and analyze trends to help artists grow their audience. [...]''
\item Task variables:
  \begin{itemize}
  \item Level: B1
  \item Age: 15--18 (teenagers)
  \end{itemize}
\end{itemize}

\noindent\textbf{Task Criteria:}
\begin{itemize}
\item Creates a complete 45-minute lesson plan (+10)
\item Activities use the provided AI/music text (+8)
\end{itemize}

\noindent\textbf{Consensus Criteria:}
\begin{itemize}
\item 02a-01: Includes appropriate pattern (PPP, ESA, TBLT) (+5)
\item 02a-02: Activities build knowledge/skills for goal (+6)
\item 02a-03: Clear structure for student profile (+8)
\item 02a-07: Realistic timings (+6)
\item 02a-09: Activities engage students (+8)
\end{itemize}

\noindent\textbf{Universal Criteria:}
\begin{itemize}
\item UC01: Language appropriateness (+2)
\item UC03: Learner profile appropriateness (+10)
\item UC04: Offensive content ($-$10)
\item UC05: Sensitive content ($-$5)
\end{itemize}

\clearpage
\section{Statistical Framework for L2-Bench Leaderboard Scores}

L2-Bench is a work in progress. We plan to release the full evaluation dataset (withholding a test set, see below) in late spring of 2026 following further construct iteration and data validation. At the same time, we will report L2-Bench evaluation results for frontier models on our dedicated website as a leaderboard and in a subsequent full paper. A hold-out set will serve as a tool to detect saturation and to update our leaderboards and dataset accordingly.

Following methodological recommendations from Miller 2024 for reliable benchmark evaluation, a model's overall L2-Bench score is an estimate with uncertainty given by its standard error. The standard error decomposes into two additive components:

\begin{enumerate}
    \item Variance of the conditional mean (super-population variance). The questions in L2-Bench do not represent all possible questions but are drawn from a hypothetical super-population of language education tasks. This component reflects uncertainty from question sampling and is irreducible - it cannot be decreased without expanding the benchmark.

    \item Mean conditional variance (response variance). Each question's score comprises a mean component (the "true" score for that question) and a zero-mean random component (response variance from stochastic generation). This component can be reduced by generating multiple responses per question and averaging.
\end{enumerate}

We generate k = 3 responses per task question and compute the mean score. This resampling strategy:

\begin{itemize}
    \item Reduces the expected conditional variance contribution to SE.
    \item Enables computation of within-question standard errors.
    \item Assumes Central Limit Theorem validity: independently drawn questions with finite variance across a sufficient number of items (N $>$ 1,000).
\end{itemize}

With this framework, alongside all point estimates we can report uncertainty via standard errors and 95\% confidence intervals. Then, when comparing models A and B, we compute question-level paired differences rather than comparing population-level summary statistics, exploiting correlation between model scores on the same questions to reduce variance. With N = 1,300 paired task comparisons at $\alpha$ = 0.05, we achieve 80\% power to detect effects as small as d $\approx$ 0.08 - sufficient to identify meaningful performance differences between frontier models where score gaps are often subtle. 

\end{document}